\documentclass[%
 aip,
preprint,
]{revtex4-1}

\usepackage{graphicx}
\usepackage{dcolumn}
\usepackage{bm}

\usepackage{amssymb,mathrsfs,amsmath}
\graphicspath{{figures/}}
\usepackage{amsmath}
\usepackage{subfigure}
\usepackage{mathrsfs}
\usepackage{subcaption} 
\usepackage{ragged2e}
\usepackage{booktabs}
\usepackage{float}
\usepackage{subcaption} 
\usepackage{comment}
\usepackage{multirow}

\usepackage{color}
\usepackage{multirow}
\usepackage {CJK}
\usepackage[utf8]{inputenc}
\usepackage[T1]{fontenc}
\usepackage{mathptmx}
\usepackage{booktabs}
\usepackage{makecell} 
\usepackage{booktabs} 
\usepackage{multirow} 
\begin{document}

\title{Using Diffusion Models for Reducing Spatiotemporal Errors of Deep Learning Based Urban Microclimate Predictions at Post-Processing Stage}

\author{Sepehrdad Tahmasebi}
\author{Geng Tian}
\affiliation{
Centre for Zero Energy Building Studies, Department of Building, Civil and Environmental Engineering, Concordia University, Montreal, H3G 1M8, Canada}%

\author{Shaoxiang Qin}
\affiliation{
Centre for Zero Energy Building Studies, Department of Building, Civil and Environmental Engineering, Concordia University, Montreal, H3G 1M8, Canada}
\affiliation{McGill University, School of Computer Science, Montreal, H3A 0E9, Canada\\}%
\author{Ahmed Marey}
\affiliation{
Centre for Zero Energy Building Studies, Department of Building, Civil and Environmental Engineering, Concordia University, Montreal, H3G 1M8, Canada}%
\author{Liangzhu (Leon) Wang}
\email{leon.wang@concordia.ca}
\affiliation{
Centre for Zero Energy Building Studies, Department of Building, Civil and Environmental Engineering, Concordia University, Montreal, H3G 1M8, Canada}%
\author{Saeed Rayegan}
\affiliation{
Centre for Zero Energy Building Studies, Department of Building, Civil and Environmental Engineering, Concordia University, Montreal, H3G 1M8, Canada}%

\begin{abstract}
-----------------------------------------------------------------------------------------------------------

\textbf{ABSTRACT}

Computational fluid dynamics (CFD) is a powerful tool for modeling turbulent flow and is commonly used for urban microclimate simulations. However, traditional CFD methods are computationally intensive, requiring substantial hardware resources for high-fidelity simulations.
Deep learning (DL) models are becoming popular as efficient alternatives as they require less computational resources to model complex non-linear interactions in fluid flow simulations. A major drawback of DL models is that they are prone to error accumulation in long-term temporal predictions, often compromising their accuracy and reliability. To address this shortcoming, this study investigates the use of a denoising diffusion probabilistic model (DDPM) as a novel post-processing technique to mitigate error propagation in DL models' sequential predictions. To address this, we employ convolutional autoencoder (CAE) and U-Net architectures to predict airflow dynamics around a cubic structure. The DDPM is then applied to the models' predictions, refining the reconstructed flow fields to better align with high-fidelity statistical results obtained from large-eddy simulations.
Results demonstrate that, although deep learning models provide significant computational advantages over traditional numerical solvers, they are susceptible to error accumulation in sequential predictions; however, utilizing DDPM as a post-processing step enhances the accuracy of DL models by up to 65\% while maintaining a 3 times speedup compared to traditional numerical solvers. These findings highlight the potential of integrating denoising diffusion probabilistic models as a transformative approach to improving the reliability and accuracy of deep learning-based urban microclimate simulations, paving the way for more efficient and scalable fluid dynamics modeling.
\end{abstract}

\maketitle

\section{\label{sec:level1} Introduction}
As urban populations have grown rapidly, urban microclimate studies have become increasingly important due to their significant impact on outdoor human thermal comfort and building energy efficiency.
Computational fluid dynamics (CFD) is crucial for simulating turbulent flows in fields such as automotive, aerospace, and environmental modeling. High-fidelity numerical approaches, such as direct numerical simulation (DNS) and large eddy simulation (LES), provide detailed solutions for complex turbulent flows \cite{article, Smagorinsky}.
While DNS offers high accuracy by simulating all turbulent scales, it is computationally expensive, especially at high Reynolds numbers \cite{ARGYROPOULOS2015693}. 
LES, which resolves large-scale eddies while modeling smaller-scale turbulence, is widely used for urban-scale simulations such as modeling airflow around buildings \cite{tian_impact_2024}.
However, LES still requires significant computational resources, as demonstrated by its 25-fold higher computational cost than Reynolds-Averaged Navier-Stokes (RANS) for simulating airflow around a cubic building \cite{tominaga2010numerical}. To mitigate the intensive computational demands of  DNS and LES methods, hybrid methods are introduced with high fidelity in predictions such as Hybrid RANS-LES models, very large eddy simulation (VLES) \cite{speziale1998turbulence}, the detached eddy simulation (DES) \cite{spalart1997comments}, and the scale adaptive simulation (SAS) \cite{menter2005scale}. 
However, these hybrid models fail to achieve the desired accuracy-cost efficiency, particularly for comparative investigations and long-term simulations with dynamic boundary conditions \cite{yi2021hybrid, MASOUMIVERKI2022108966}.

To address the trade-off between computational demand and simulation precision in traditional numerical solvers, data-driven methods have introduced deep learning (DL) as an efficient alternative to solve partial differential equations (PDEs). Convolutional neural networks (CNNs), particularly U-Nets and convolutional autoencoders (CAEs), perform well in predicting flow fields by capturing spatially complex features from high-dimensional data \cite{mohan2018}.
Reduced-order models (ROMs) simplify high-dimensional datasets into manageable forms, enabling applications such as flow prediction around cylindrical bodies \cite{LIBERGE2010292, Rajetall}, airfoil dynamics \cite{act13030088}, and urban wind flow modeling \cite{XIANG2021107397}. As specialized ROMs, CAEs provide non-linear dimensionality reduction by encoding data into a compact latent space, where sequence modeling networks such as long-short-term memory (LSTM) networks \cite{LSTM} predict temporal flow evolution. This framework has been applied to turbulent channel flows \cite{Nakamura2021} and urban flows around high-rise structures \cite{MASOUMIVERKI2022104252}.
U-Nets, having CNN strengths, are increasingly adopted in CFD for retaining localization through their downscaling and upscaling pathways. They have been used to predict velocity and pressure in 2D laminar flows \cite{chen}, hydrodynamic flows for submarines \cite{hou2022novel}, and vortex detection in complex flows \cite{DENG2022108229}, proving their robustness for high-resolution fluid dynamics applications \cite{ronneberger2015, Shelhamer}. DL models have made significant strides in accuracy and efficiency compared to traditional numerical solvers. However, predicting flow fields over long time horizons remains challenging \cite{NEURIPS2020_43e4e6a6}. The chaotic nature of turbulence can make the predictions even more uncertain, as small initial deviations from the initial spatial state can lead to large errors in predictions over time, making these models highly sensitive to initial conditions \cite{Pope_2000}.
Research has consistently recommended improving initial prediction accuracy in DL models to control error propagation across successive time step predictions \cite{halder2024reduced, hasegawa2020machine}. Despite improvements in initial prediction accuracy, DL models often fail to fully capture the complex statistical dynamics of turbulence, limiting their ability to generate high-fidelity flow fields over long-time predictions \cite{chen2023fuxi}. To address this issue, data assimilation techniques, physics-informed neural networks (PINNs), and generative models have been widely utilized in the literature to reduce the error accumulation in DL models. 

Data assimilation techniques are increasingly used with machine learning to reduce the error accumulation in long-term predictions and better predict turbulent flow dynamics \cite{10141545, casas2020reduced}. By integrating observed data into DL models, these techniques continuously adjust predictions, enhancing accuracy and reliability in representing the system's state \cite{BUIZZA2022101525}. Methods such as ensemble Kalman inversion have shown effectiveness in modeling separated flows at high Reynolds numbers and capturing prominent turbulent features \cite{wang2022}. Other advancements include modifying the Spalart-Allmaras turbulence model with spatial variation coefficients for adaptability \cite{Chuangxin} and optimizing sensor placement using deep neural networks within ensemble Kalman filters \cite{deng2021deep}. Despite these advancements, data assimilation relies largely on rich datasets from CFD simulations and experiments, presenting challenges in creating generalizable models for diverse turbulent flow scenarios \cite{wang2022}.
Although this reliance on extensive data sources poses challenges in scaling and adapting these methods to diverse applications, the research continues the refinement of data assimilation frameworks.

PINNs are a class of DL models designed to solve the physical system by integrating the physical laws into the learning phase of the DL model for PDEs calculations, which initially was introduced to solve the 1D Burgers equation and later applied to solve the 2D Navier-Stokes equations for a flow past a cylinder \cite{RAISSI2019686}. In the PINNs framework, knowing the governing PDE equations of the system allows the neural network to train to minimize the residuals of these equations, which ensures that the network's predictions not only match the data but also adhere to the underlying physical laws. Applications of PINNs have been widely studied in fluid dynamic problems, such as simulating the vortex-induced vibrations \cite{raissi2019deep}, flow over a cylinder \cite{ren2024physics}, and turbulent channel flow \cite{jin2021nsfnets}. Although PINNs have shown promising improvement in long-term predictions compared to traditional DL models \cite{sharma2023review}, their application remains limited to case studies where the system's governing PDEs are known.

In response to the challenges of accurately predicting turbulent flow fields, generative models have emerged as robust solutions, effectively capturing the complex stochastic behavior of turbulence \cite{GAO2024117023}. These models, built on advanced neural network architectures, learn high-dimensional probability distributions from finite datasets and synthesize new samples that maintain the statistical properties of the original data \cite{Ruthotto}.
Variational autoencoders (VAEs) \cite{kingma2013auto} and generative adversarial networks (GANs) \cite{goodfellow2014generative} have shown promising potential in fluid dynamics applications. VAEs utilize a probabilistic approach to overcome the drawbacks of traditional autoencoders, offering smoother latent spaces for generating realistic new instances. This approach has been applied to various tasks, including airflow predictions around airfoils \cite{wang2021flow}, cylinders \cite{wang2024towards}, and the generation of turbulent flow fields \cite{grogan2017variational}.
GANs have a complementary approach, employing adversarial training to learn probability distributions and generate synthetic flow fields that closely resemble real data. They are capable of tasks such as super-resolution of turbulent velocity fields \cite{deng2019super, yu2022three} and modeling complex turbulent flow patterns \cite{buzzicotti2021reconstruction, drygala2022generative}. These advancements make both VAEs and GANs valuable tools for modeling and predicting fluid dynamics.

Diffusion models \cite{Jascha} have recently developed as an effective class of deep generative models, achieving high performance across different fields, including image synthesis \cite{dhariwal2021diffusion, song2020score}, computer vision \cite{zimmermann2021, zhao2022egsde}, and natural language processing \cite{savinov2022, yu2023}. These models use a two-stage process: (I) adding noise to disrupt data structure in the forward step; and (II) iteratively denoising the data with a neural network to reconstruct the original data.
Diffusion models encompass several types, including denoising diffusion probabilistic models (DDPMs) \cite{ho2020}, score-based generative models (SGMs) \cite{song2019}, and stochastic differential equations (SDEs) \cite{song2020score}. 
Diffusion models offer key advantages over traditional generative models like VAEs and GANs. They feature simpler training with minimal hyperparameter tuning \cite{GAO2024117023}, maintain the same dimensionality between latent variables and input data for better high-frequency detail capture \cite{kingma2013auto}, and avoid the convergence issues common with GANs, ensuring greater stability and robustness \cite{metz2017}.
While widely used in computer vision and super-resolution, diffusion models remain underexplored in fluid dynamics, though recent studies are beginning to uncover their potential.
For instance, diffusion models have been used for super-resolving low-fidelity flow fields \cite{SHU2023111972} and reconstructing turbulent flows \cite{kohl2024, lippe2023pde} by integrating diffusion processes with physics-informed neural networks (PINNs). These approaches, leveraging simplified 2D PDEs, have effectively enhanced low-resolution flow data and captured complex flow dynamics \cite{RAISSI2019686}.
Another study showed that DDPMs can learn the underlying properties of various turbulent flow states, generating velocity fields, independently of specific initial conditions \cite{lienen2024}. This generalization offers a key advantage over traditional autoregressive models, which often suffer from error accumulation when generating flow fields from initial conditions.
While physics-informed diffusion models are effective for reconstructing turbulent flows, they depend on known governing PDEs. In urban microclimate studies, CFD simulations are typically performed in 3D domains, but DL models are often trained on 2D planes to reduce computational costs \cite{MASOUMIVERKI2022104252, HU2024111120}. This dimensional reduction leaves the governing PDEs for the selected planes undefined, posing a challenge to their integration.

The above-mentioned shortcomings motivate this study to explore a new approach for reducing the error accumulation of DL models, aiming to generate more accurate flow fields in urban microclimate predictions.
Specifically, our study examines the application of 
DDPMs as a post-processing technique to mitigate spatial discretization errors in DL models for urban microclimate predictions, eliminating the need for experimental data or prior knowledge of the flow field's underlying physics.
Most existing studies rely on data assimilation or PINNs to reduce error accumulation in multi-step predictions, but their use in urban microclimate simulations is limited by a lack of experimental data for different test cases as well as unknown PDEs.
In addition, our study further conducts a comparative analysis of the one-step and sequential time-step prediction performance of two prominent architectures, the CAE and U-Net, in simulating turbulent airflow around a building block. The other contribution of our study is to measure the advantage of integrating DDPM post-processing on the reconstructed images generated by these models, by evaluating the enhancements in the fidelity of the turbulence simulations.
Overall, our paper contributes to understanding the potential of DDPM in improving spatial resolution and accuracy in DL-based fluid dynamics applications.

The remainder of the paper continues as follows: Section \ref{method} describes the proposed framework, the case study, and the DL models used for flow field reconstruction. Section \ref{sec_RAD} focuses on the CFD simulation used to generate the dataset, validation of the results, description of the dataset used for training the DL model,  the results of the DL model, and the effectiveness of the proposed framework in sequential time step predictions. Finally, Section \ref{sec.conclusion} presents the conclusions and potential directions for future work.

\section{Methodology}
\label{method}
\subsection{Implementation of DDPM}
The denoising diffusion probabilistic model (DDPM) leverages the Markov chain property to generate data by sampling from Gaussian distributions. This generative model is widely used in computer vision, natural language processing, and numerous other areas \cite{Haoying,ozbey2023unsupervised, Li}. The mathematical details concerning DDPM are provided in Appendix \ref{sec.appendixA}. The diffusion model includes forward and backward processes as shown in Figure \ref{Figureddpm}. 
The DDPM framework has two hyperparameters, the number of steps $N$ and the noise schedule $\beta_t$, which determine the rate at which data samples are converted into Gaussian noise. In the present study, $N = 1000$ and a linear noise scheduler from $\beta_1 = 10^{-4}$ to $\beta_T = 0.02$ is utilized \cite{ho2020}. In addition, the model employs a U-Net architecture with residual blocks, attention mechanisms, and positional embeddings to estimate $q(x_{n-1}|x_0, x_n)$ \cite{ho2020}. This architecture features 4 levels of downsampling, each by a factor of 2.

\begin {figure}[h!]
	\centering{\includegraphics[scale=0.48]{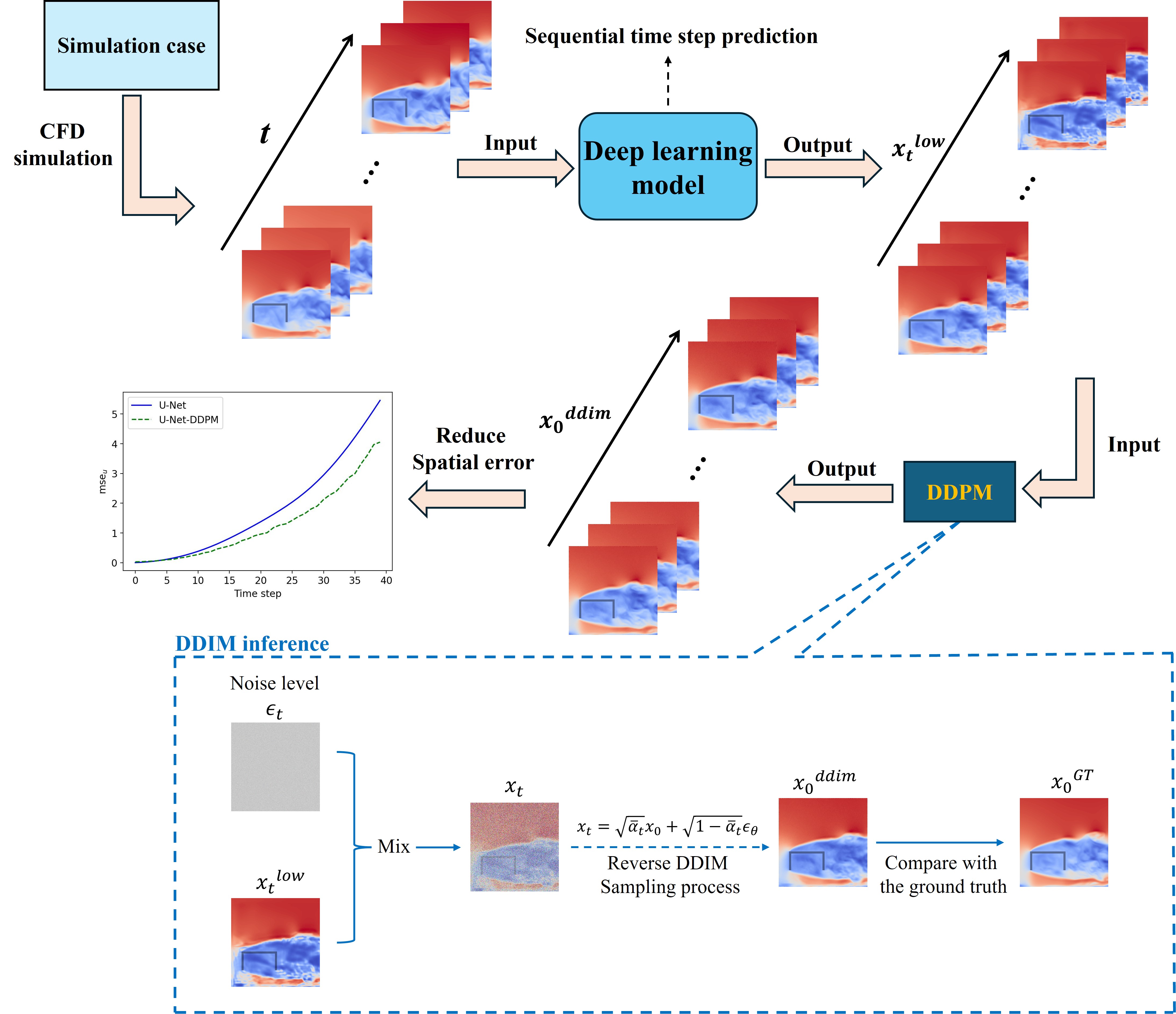}}
	\caption{An overview of the proposed framework for reconstructing high-fidelity flow fields from low-quality images from DL models using DDPM with DDIM sampling.}
	\label{Figureddim}
	\end{figure}

As the DDPMs require simulation of Markov chains for many steps in order to generate a sample, denoising diffusion implicit model (DDIM) are introduced to accelerate the sampling process of DDPM.
DDIM modifies the reverse diffusion process to enable deterministic sampling, allowing the generation of \( x_0 \) directly from \( x_t \) at any chosen time step. This deterministic characteristic eliminates the stochasticity inherent in DDPM sampling, offering faster generation while preserving high sample quality \cite{Jiaming}. 
In other words, DDPM is capable of being trained over an arbitrary number of forward steps, but it can sample from some of those steps in the generative process using DDIM. The prediction of the DDPM model and any step using DDIM sampling can be described as follows: 

\begin{equation}
x_t = \sqrt{\bar{\alpha}_t} x_0 + \sqrt{1 - \bar{\alpha}_t} \epsilon_\theta
\end{equation}

where \( x_0 \) is the high quality data sample, \( x_t \) represents the noisy data at time-step \( t \), \(\epsilon_{\theta}\) is the predicted noise by the model, and \( \bar{\alpha}_t \) is a hyperparameter obtained from Equation \ref{ddpm-equ}.
Figure \ref{Figureddim} demonstrates the inference of the proposed framework.

Initially, high-fidelity CFD data sequences are processed by a pre-trained DL model to generate a turbulent flow field by sequential time step predictions. Due to the error accumulation during the sequential predictions, low-fidelity flow fields will be generated. 
low-fidelity images are then input to a pre-trained DDPM as a post-processing step to reduce spatial discrepancies and enhance image fidelity.
During the training part of the DDPM, the model learns to make the predictions of \(x_0\) (synthetic flow field) using the starting point of \(x_T\) (fully noisy flow field) so that \(x_0\) closely approximates the ground truth \(x_0^{GT}\) (high-fidelity CFD data). 
In the model inference, given a low-fidelity flow field predicted by a DL model, denoted as \(x_t^{low}\), is refined using a reverse diffusion process with DDIM sampling.
The low-fidelity flow field is mixed with the random noise \(\epsilon_t\), and the model aims to generate a synthetic flow field \(x_0^{ddim}\) so that it minimizes the discrepancy from the ground truth \(x_0^{GT}\). By adopting this approach, the model can predict flow fields over extended time steps while maintaining acceptable accuracy. 
Consequently, in the reverse diffusion process, the initial sampling step—which determines the level of injected noise—is a crucial factor. In addition, the availability of accurate data within the low-fidelity image is another important factor, as DDPM uses sparse samples within the low-fidelity flow field to reconstruct an authentic image that more accurately reflects the underlying dynamics of the fluid flow \cite{SHU2023111972}.
Notably, since DDIM sampling is utilized, the number of reverse process sampling steps is set to 25 to effectively balance computational efficiency and accuracy.

\subsection{Simulation Case}
\begin{figure}[h!]
    \centering
    \subfigure[]{
        \includegraphics[width=0.85\linewidth]{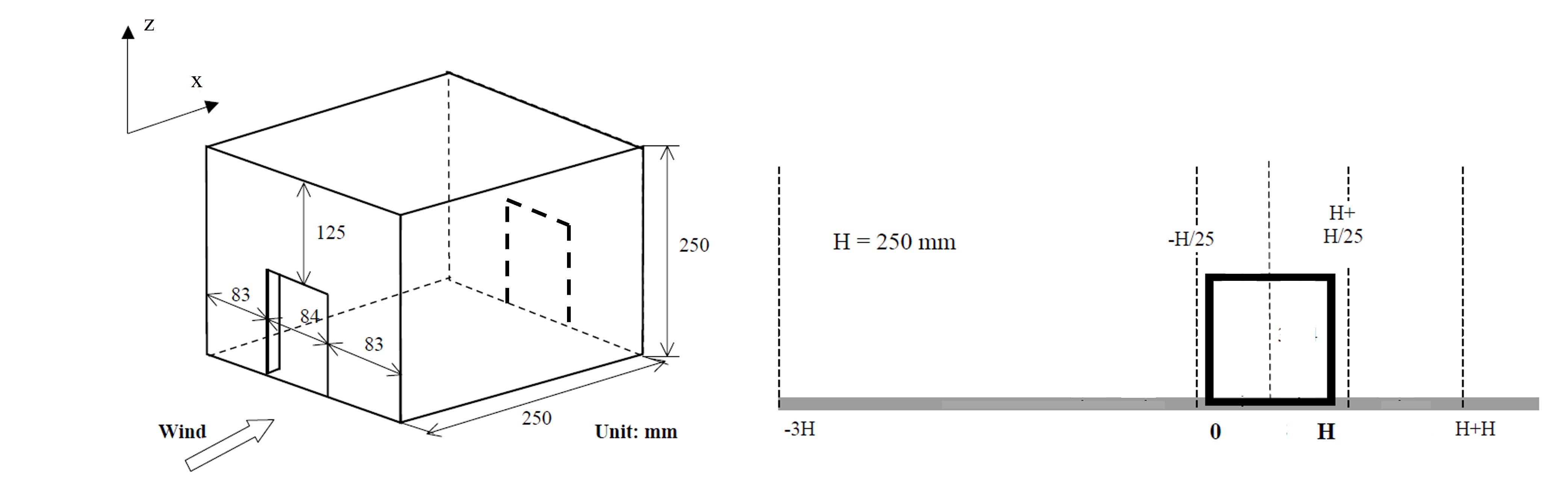}
        \label{fig:schematic}
    }

    \subfigure[]{
        \includegraphics[width=0.7\linewidth]{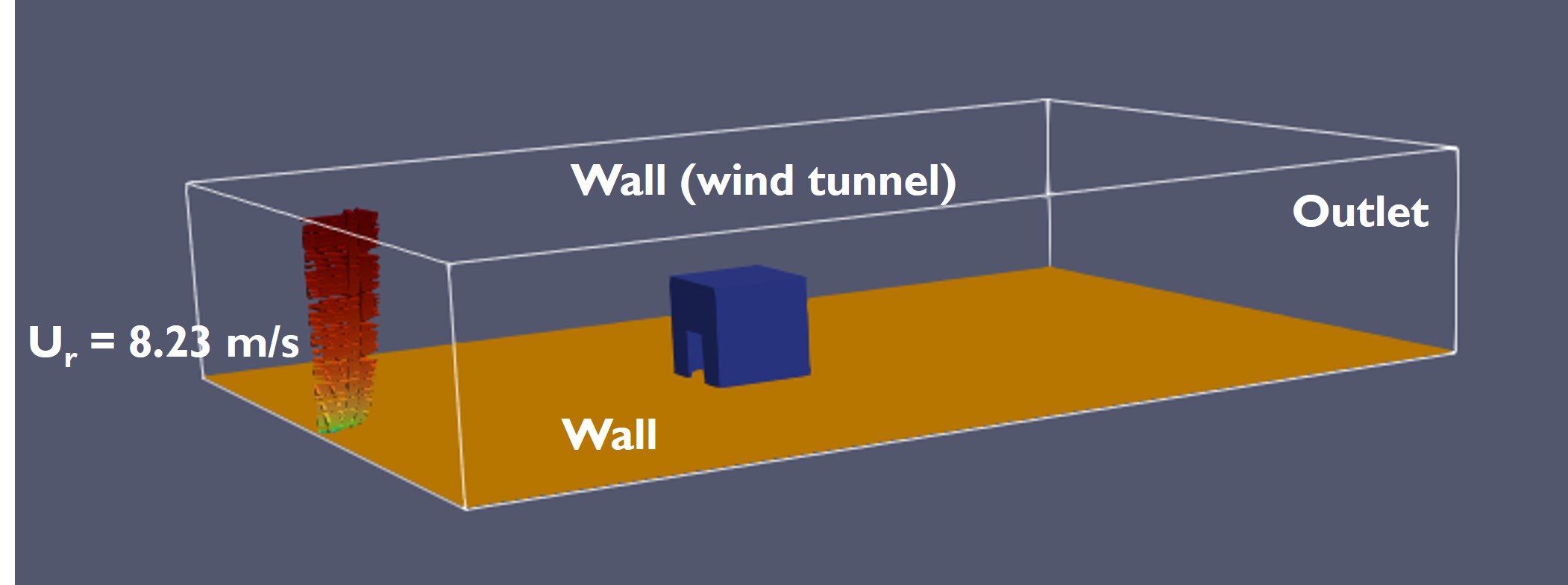}
        \label{fig:simulation domain}
    }

    \caption{(a) A schematic of the cubic structure \cite{JIANG2003331}; (b) Simulation domain}
    \label{fig_view}
\end{figure}

The current simulated case is a small cubic structure with two openings in the windward and leeward direction of the structure, first carried out experimentally in a boundary layer wind tunnel under highly turbulent flow \cite{JIANG2003331}. The cube's height, width, and depth are $250\ mm$, $250\ mm$ and $250\ mm$, respectively. The height and width of the openings are $125\ mm$ and $84\ mm$, respectively, and the thickness of the wall is $6\ mm$. The Reynolds number is set to 140,000, and the inflow velocity profile gives $\frac{U^*}{U_r} = \left(\frac{h^*}{H_{\text{ref}}}\right)^{0.17}$.  The schematic of the cube and different locations where air velocity was measured in the experiment are depicted in Figure \ref{fig:schematic}, and the simulation domain along with the boundary conditions are depicted in Figure \ref{fig:simulation domain}.

The training and testing datasets are generated from the CFD simulation conducted by CityFFD \cite{mortezazadeh2019cityffd}. Using the semi-Lagrangian approach and fractional stepping method, CityFFD is able to simulate large-scale urban microclimate using GPU computational acceleration scheme \cite{MORTEZAZADEH2020106955}. More information about CityFFD mathematics and workflow can be found in previous works \cite{MORTEZAZADEH2022101063, Mohammad, Mortezazadeh2017}. CityFFD has previously been applied to modeling urban wind flow \cite{Jiwei}, extreme weather conditions through integration with energy models \cite{KATAL20191402}, and modeling urban microclimate during heatwaves by integration with Weather Research and Forecasting model \cite{MORTEZAZADEH2021102670}. 

The LES solver in CityFFD solves the following non-dimensional Navier-Stokes equation:
\begin{equation}
\begin{aligned}
\nabla \cdot \vec{U} &= 0 \\
\frac{\partial \vec{U}}{\partial t} + (\vec{U} \cdot \nabla) \vec{U} &= -\nabla p + \left(\frac{1}{Re} + v_t\right)\nabla^2 \vec{U} - \frac{Gr}{Re^2}T \\
\frac{\partial T}{\partial t} + (\vec{U} \cdot \nabla)T &= \left(\frac{1}{Re \cdot Pr} + \alpha_t\right)\nabla^2 T + q
\end{aligned}
\end{equation}
where $U$, $t$, $p$, $v_t$, $Re$, $Gr$, $Pr$, $T$, $q$, and $\alpha_t$ are dimensionless velocity, time, pressure, turbulent viscosity, Reynolds number, Grashof number, Prandtl number, temperature, heat source, and turbulent thermal diffusivity, respectively. 

The Smagorinsky subgrid-scale model used here is based on the following equation:
\begin{equation}
v_t = \left(c_s \Delta \right)^2 |\overline{S}|
\end{equation}
where $c_s$, $\Delta$, and $\overline{S}$ are Smagorinsky constant, the filter width, and the strain rate, respectively. The Smagorinsky constant used in the current study is 0.2, which is in the range of $0.1 \leq c_s \leq 0.24$ suggested for wall-bounded turbulent flow.

\subsection{Deep Learning (DL) Models Description}
\label{sec.DLs}

Once high-fidelity CFD results are generated, they serve as training data for DL models to predict airflow around a cubic structure. After training, these models can be deployed to forecast airflow around the target building.

This study focuses on two widely used DL architectures in fluid dynamics: CAE-LSTM and U-Net, which are employed to reconstruct the flow field. The detailed configurations and structures of these models are provided in Appendix \ref{sec.appendixA}. Below, we outline the key features and configurations of each model in this context.

\subsubsection{CAE-LSTM}

Convolutional autoencoders (CAEs) combine convolutional neural networks (CNNs) with autoencoder structures to efficiently handle spatially structured data like images. The encoder compacts the data into compressed latent space using convolutional layers, while the decoder reconstructs the original state, optimizing reconstruction quality through loss functions like mean squared error (MSE). This makes CAEs effective for tasks such as image denoising, data reconstruction, and unsupervised learning \cite{Jiang, FANG201896, Naderan, zhu2021stacked}. 
The structure of the CAE network is shown in Figure \ref{Figureautoencoder}.
In this study, downsampling blocks are designed with convolutional layers using strides larger than one to reduce spatial dimensionality while retaining important features \cite{zhu2021stacked}. The upsampling blocks in the decoder use transpose convolutions to incrementally bring the data back to its initial resolution. To more effectively capture temporal dependencies in data sequences, the CAE architecture is integrated with a parallel Long Short-Term Memory (LSTM) network, shown in Figure \ref{Figureparallel LSTM}. The LSTM network efficiently uses information from the preceding 20 time steps (detailed results not shown here). This configuration, which balances accuracy and computational efficiency, enhances the model’s capability to process spatiotemporal data \cite{MASOUMIVERKI2022104252}.

\subsubsection{U-Net }
$\textbf{U-Net}$ is a convolutional neural network primarily designed for biomedical image segmentation, and presents an efficient alternative to the traditional sliding window approach used in conventional convolutional networks \cite{ronneberger2015, ciresan2012deep}.
An overview of the $\textbf{U-Net}$ architecture employed in the present study is illustrated in Figure \ref{Figureunet}.
The $\textbf{U-Net}$ architecture includes a down-sampling phase to halve the image size and double the channels, and an up-sampling phase to restore the size. In addition, skip connections between corresponding layers preserve localization information.
This study explores two variants of the $\textbf{U-Net}$ architecture for predictive modeling in fluid dynamics. The first model is trained to reconstruct the next frame based solely on the most recent observation, while the second model, referred to as $\textbf{U-Net20}$, incorporates the prior 20 time steps to capture temporal dependencies (see Table \ref{tab:unet20}). $\textbf{U-Net20}$ is designed to align with the CAE-LSTM model, which also utilizes the last 20 observations, enabling a direct comparison of their abilities to generate flow fields from equivalent temporal inputs.

\subsection{Models training}
This section explores the training process of CAE-LSTM, U-Net models, and DDPM. Table \ref{tab_training_hyperparameters} outlines the hyperparameters applied across these DL models. Both CAE and U-Net architectures utilize the Rectified Linear Unit (ReLU) activation function \cite{nair2010rectified} and are trained to minimize mean squared error (MSE) as the loss function. However, the DDPM employs Sigmoid-Weighted Linear Units (SiLU) \cite{elfwing2018sigmoid} to optimize noise learning dynamics, with a loss function based on the squared difference between predicted and actual noise during the forward diffusion process (see Equation \ref{eq:simple_loss}). 
As a particular form of the Swish function, the SiLU activation function has a smooth, nonlinear function with a continuous derivative, which helps prevent sudden gradient changes and allows stable training in diffusion models.
In addition, to prevent overfitting, an early stopping algorithm with a patience threshold of 20 epochs is employed for CAE, LSTM, and U-Net training, while a 10\% dropout rate is applied in DDPM.

\begin{table}[h!]
    \centering
    \caption{Hyperparameters used for training the models.}
    \begin{tabular*}{\textwidth}{@{\extracolsep{\fill}} lccccc}
        \hline
        Model & Activation function & Optimizer & Learning rate & Batch size & Epochs \\
        \hline
        CAE      & ReLU & ADAM & 0.0001 & 32 & 200 \\
        LSTM     & Tanh & ADAM & 0.001  & 32 & 250 \\
        U-Net    & ReLU & ADAM & 0.001  & 32 & 100 \\
        U-Net20  & ReLU & ADAM & 0.001  & 32 & 250 \\
        DDPM     & SiLU & ADAM & 0.0001 & 32 & 300 \\
        \hline
    \end{tabular*}
    \label{tab_training_hyperparameters}
\end{table}

\section{Results and discussion}
\label{sec_RAD}

This section evaluates the effectiveness of the proposed DDPM framework in reducing error accumulation in DL models. Initially, the CityFFD data is validated against experimental data and previous CFD simulations. Following this, the dimensionality reduction capabilities of the CAE model and the one-step prediction accuracy of the DL models used in this study are assessed, with detailed analyses provided in Appendix \ref{sec.quadrantB}. The performance of DL models in sequential time step predictions is also examined to provide a baseline for further analysis. Finally, the performance of DDPM as a post-processing technique for reducing accumulated errors in DL models is evaluated. 

\subsection{CFD Validation and Dataset}

\begin {figure}[h!]
	\centering{\includegraphics[scale=0.55]{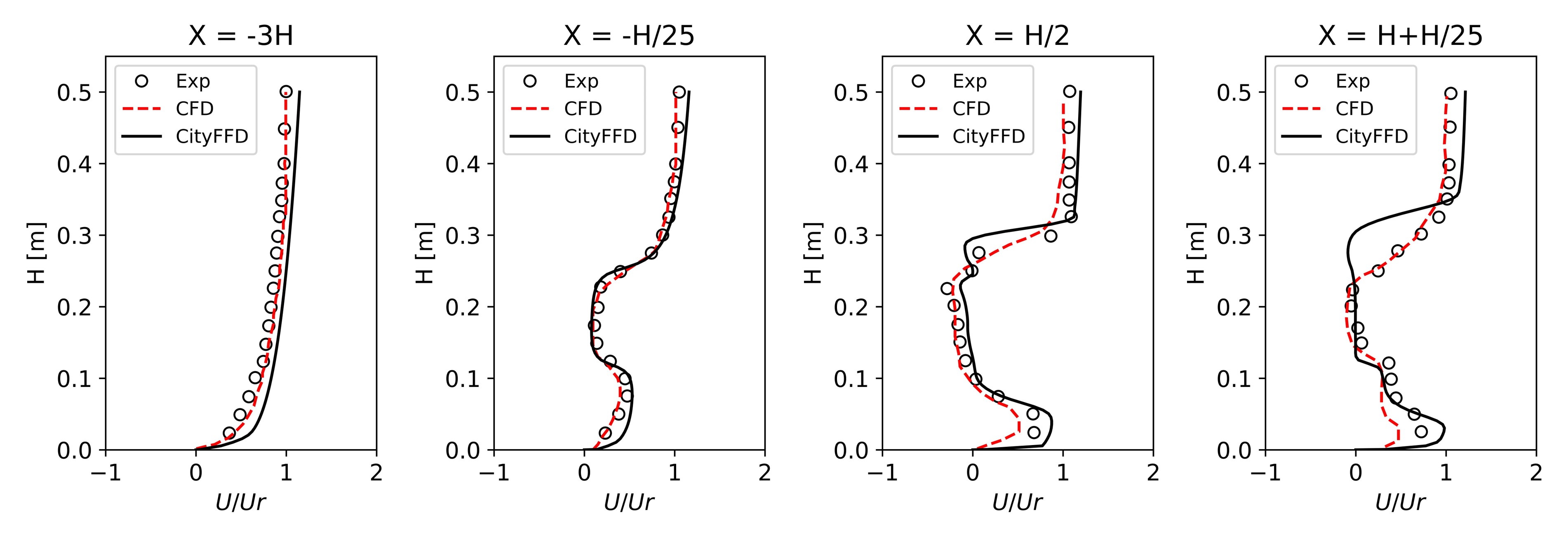}}
	\caption{Comparison of the mean velocity profile over a cubic building between the previous experimental results \cite{CHENG2018160},  numerical data \cite{JIANG2003331} and current CityFFD data. }
	\label{Figurecomparsion}
	\end{figure}

This study evaluates the performance of CityFFD in simulating airflow around a cubic structure, comparing the results with the prior CFD simulations \cite{CHENG2018160} and experimental data \cite{JIANG2003331}.
Figure~\ref{Figurecomparsion} shows the comparison of mean velocity profiles between the current results and the experimental data. At locations, $X = -3H$ and $X = -H/25$, CityFFD results are consistent with the experimental data. At the leeward side of the structure, while below the structure height, the results show close alignment with the experiment data. 
However, for the position of $X=H + H/25$, the simulation slightly underestimated the velocity compared to the experimental measurements above the structure.
This underestimation is attributed to the wall effects on the airflow pattern, as the CityFFD software lacks wall functions near the walls, resulting in a linear velocity gradient profile near the roof surface \cite{MORTEZAZADEH2022101063}. 

\begin{figure}[h!]
    \centering
    \subfigure[]{
        \includegraphics[width=0.38\linewidth]{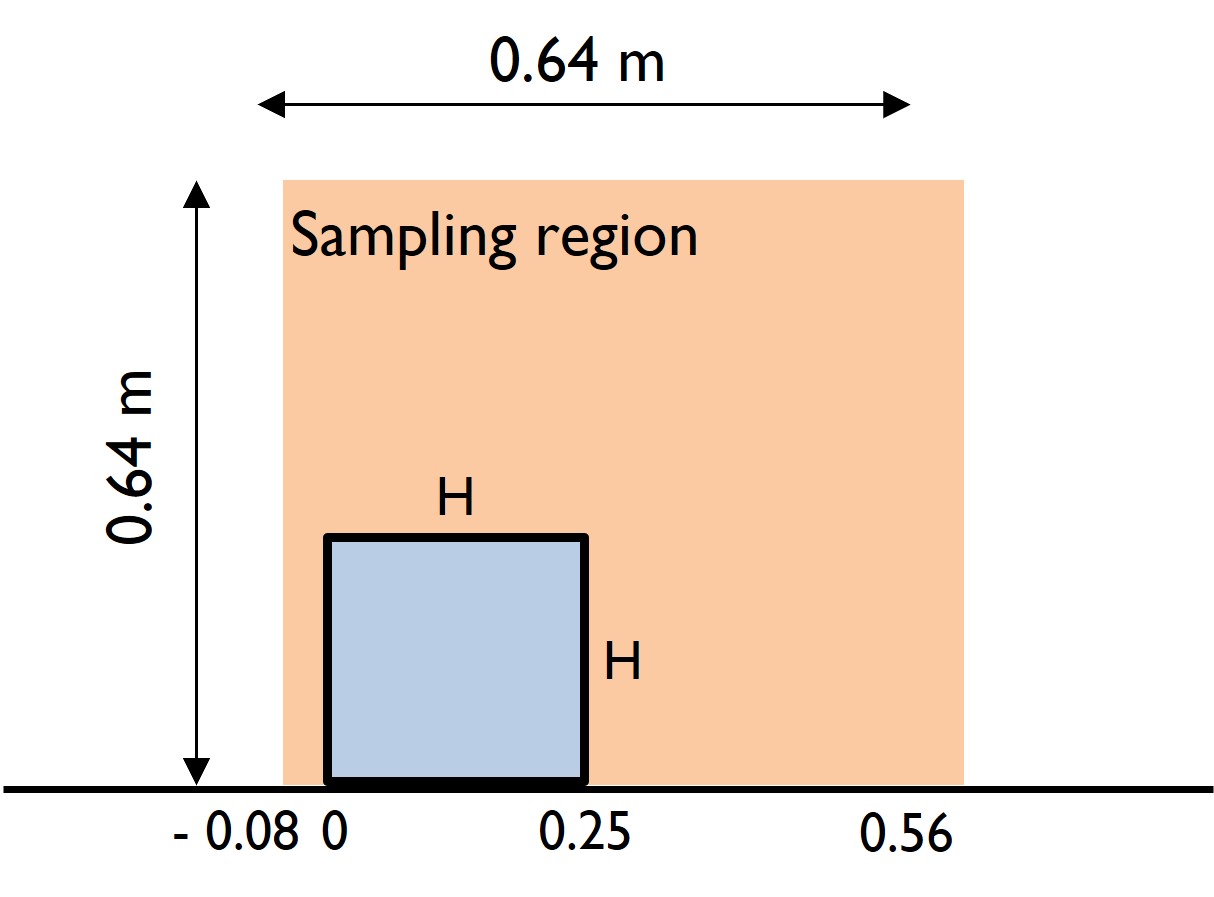}
        \label{fig:sampling region}
    }
    \hspace{0.5cm}
    \subfigure[]{
        \includegraphics[width=0.28\linewidth]{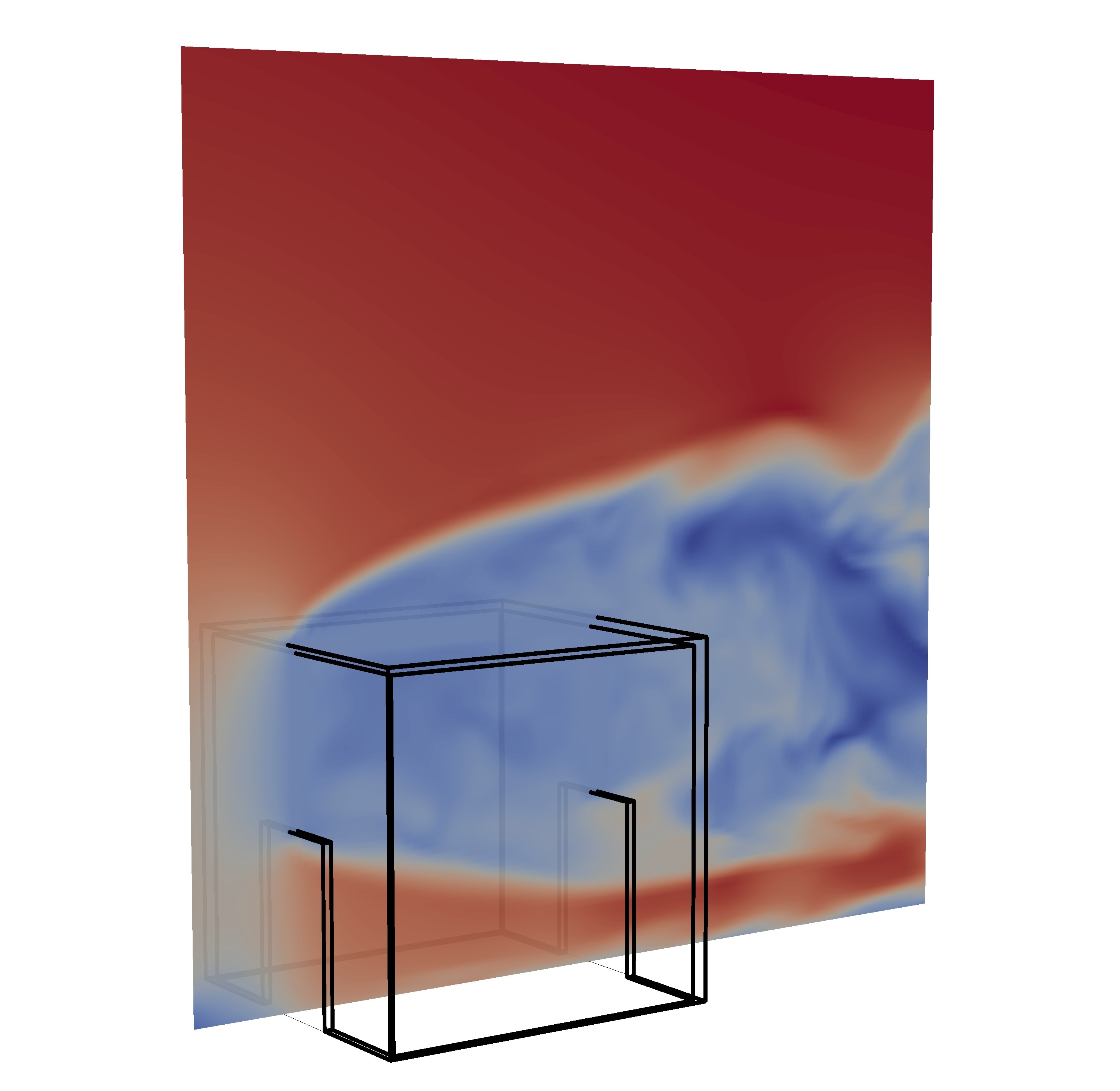}
        \label{fig:sampling contour}
    }

    \caption{(a) Data sampling region; (b) Velocity contour.}
    \label{fig_second_row}
\end{figure}

This study focuses on the longitudinal ($u$) velocity component for training CAE-LSTM, U-Net, and DDPM in flow field reconstruction. Initially, both longitudinal ($u$) and vertical ($w$) velocity components are extracted from a large-eddy simulation dataset spanning 12,000 time steps (7.2 seconds of physical time) with a flow turnover time of 0.4 seconds. However, due to the low distribution of $w$ velocity caused by high turbulence in the streamwise direction, only the $u$ component is used for model training (see Figure \ref{Figuredistribution}). 
To ensure a fully developed flow state, the first 4,000 time steps, marked by instability, are excluded, leaving 8,000 time steps (12 flow turnovers) for training and evaluation. Given the high GPU memory demands of handling velocity magnitude data across a grid of 2.3 million cells, data collection is optimized by focusing on a region around the building, which includes both the wake region and measurement points. As a result, the velocity components are projected onto $x-z$ plane intersecting the center of the cubic structure (Figures. \ref{fig:sampling region} and \ref{fig:sampling contour}), yielding a 128 × 128 grid with 5 mm resolution. Each dataset sample has dimensions of 128 × 128 × 1, representing the $u$ and $w$ velocity components, respectively. The dataset is split into training and testing subsets, with 80\% used for training and 20\% set aside for testing. This selection allows efficient handling of high-resolution flow data, facilitating accurate model training and testing for fluid dynamics predictions.

\subsection{Baseline Performance: Sequential Time Step Prediction of DL Models}

Following the examination of the predictive capabilities of CAE and U-Net models (see Appendix \ref{sec.quadrantB}), this section considers their utility in sequential time step prediction applications and builds on their promising accuracy demonstrated in one-step predictions (see Table \ref{table.pred-error}). When using the prediction results for further reconstructions of the flow field, errors tend to accumulate at each step of sequential prediction. As a result, the application of DDPM to lower the accumulated error is studied.

To generate new samples, the last observations of the ground truth are taken, and then the model is applied sequentially 40 times. A distance of 40 steps is identified as the point where two steps in the dataset are nearly uncorrelated. Subsequently, these outputs are used as inputs for the DDPM, which mitigates the spatial discrepancies due to the accumulation of errors from multiple prediction rollouts.  
In the forward diffusion process, noise is incrementally added to the image at every time step to eventually result in a fully noisy image. This process is similar to the gradual accumulation of errors observed at each stage of sequential time step prediction. As a result,
the reverse diffusion process mitigates spatial errors within the prediction outputs by conceptualizing denoising as the progressive refinement of the flow field toward the ground truth distribution. Therefore, it is possible to enhance the overall fidelity of a sequential prediction framework by using the DDPM. 

\begin {figure}[h!]
	\centering{\includegraphics[scale=0.6]{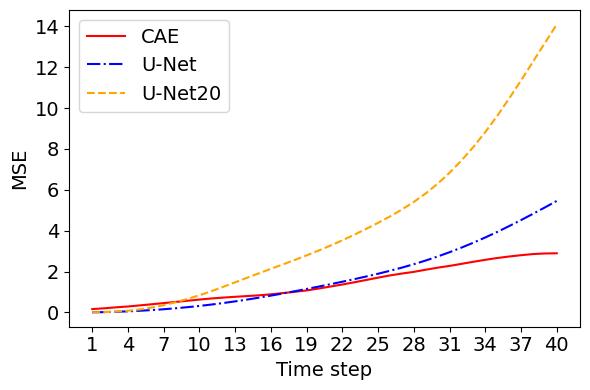}}
	\caption{Mean squared error, $MSE$, of different models over 40 steps of sequential time step prediction compared to the CityFFD results for the longitudinal velocity components.}
	\label{Figureforecasting_u}
	\end{figure}

Figure \ref{Figureforecasting_u} depicts the mean squared error of each step of the DL model compared to the ground truth.

Despite the high prediction accuracy achieved by each model in one-step predictions (see Table \ref{table.pred-error}), substantial error accumulation is observed during sequential time step predictions.
This underscores the need for an alternative and more effective framework, such as the application of DDPM in this case, to reduce error accumulation during sequential time step prediction.
In both one-step and sequential time-step prediction applications, U-Net performs better than U-Net20. As a result of the model's increased dependence on wide historical time series information, U-Net20 exhibits an accelerated divergence.
Although leveraging multiple past observations can enhance the model's understanding of long-term dependencies, it can also increase its sensitivity to accumulating errors.
Therefore, the U-Net20 model demonstrates noticeable growth in reconstruction error as inaccurate predictions accumulate and amplify. 
Compared to U-Net models, CAE exhibits a relatively lower error. This performance is primarily due to the decoder part of the CAE. The decoder, which operates as a generative model, aims to maintain the output within the distribution described by the training dataset \cite{kingma2013auto}. Consequently, the reconstructed flow fields of the CAE model have lower error values compared to other models. 
Despite the relatively low errors, the results may not fully reflect a high level of accuracy in capturing complex flow characteristics. Therefore, the CAE outputs remain within the training data distribution and may not indicate that the predicted velocity component accurately reflects the flow field dynamics.

\subsection{DDPM Post-Processing Framework Analysis}

\subsubsection{Sequential Prediction Error Reduction (Focusing on the spatially averaged MSE)}

\begin{figure}[h!]
    \centering
    \subfigure[CAE-LSTM]{
        \includegraphics[width=0.48\linewidth]{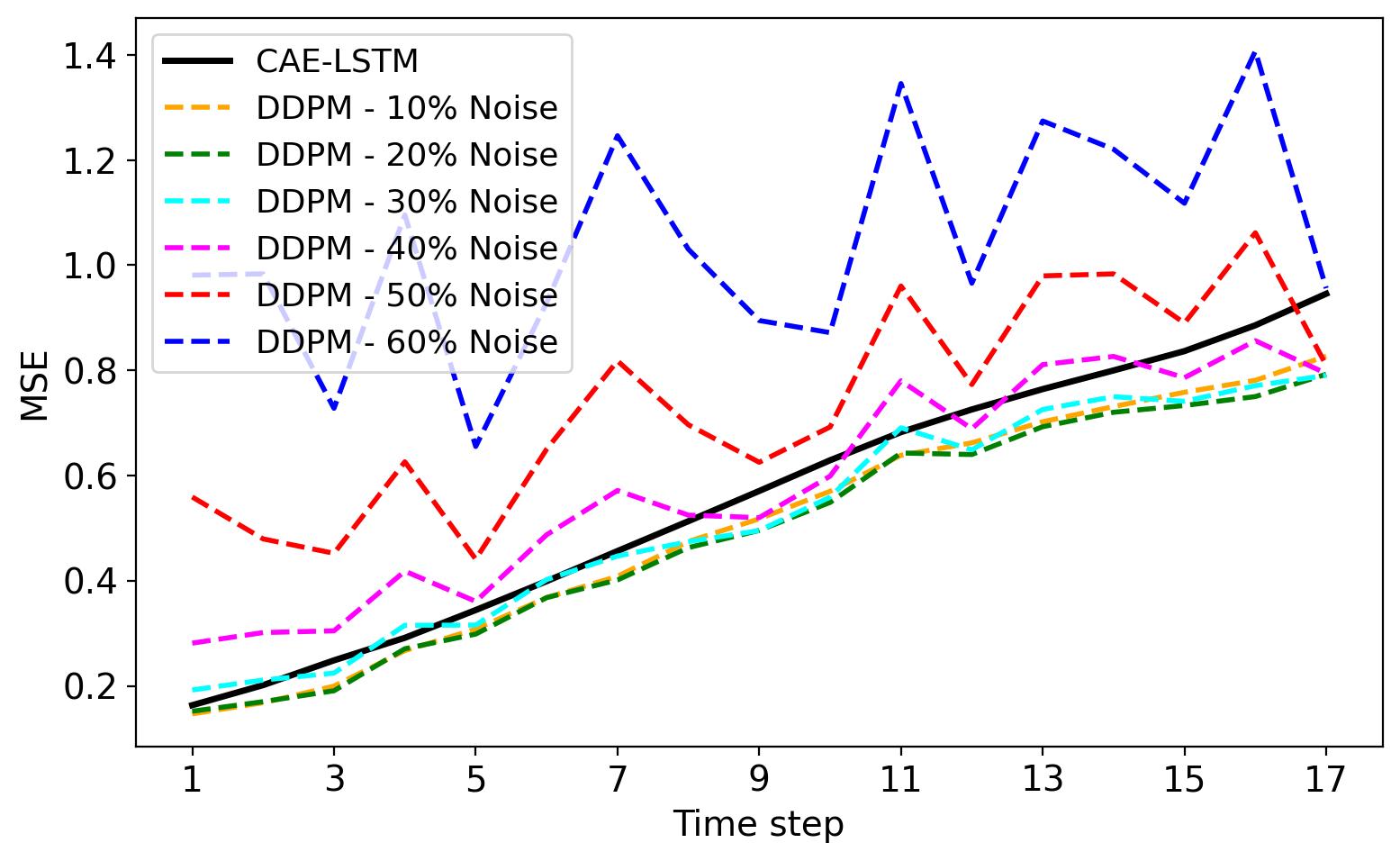}
        \label{fig:u-cae}
    }

    \subfigure[U-Net]{
        \includegraphics[width=0.48\linewidth]{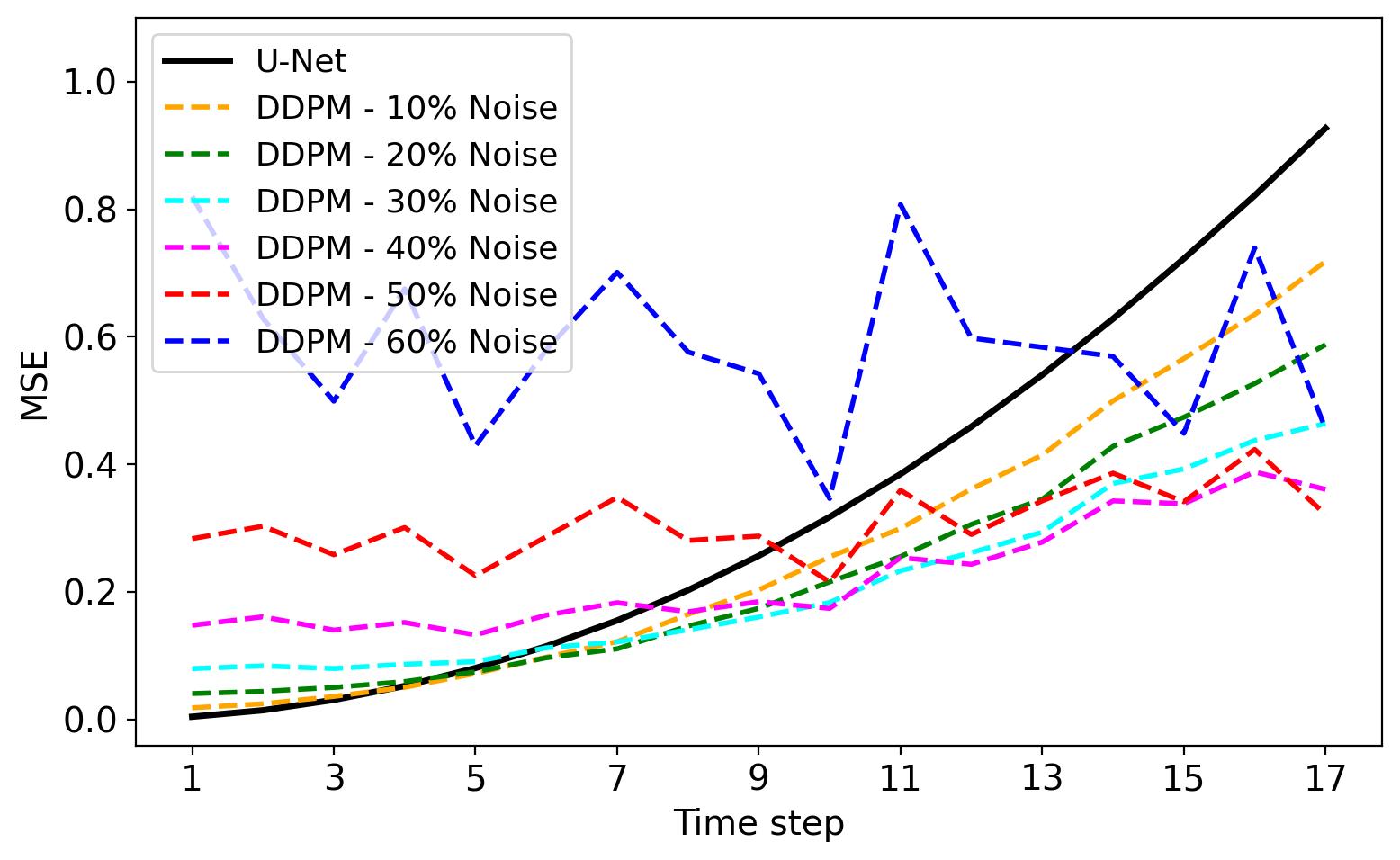}
        \label{fig:u-unet}
    }
    \hfill
    \subfigure[U-Net20]{
        \includegraphics[width=0.48\linewidth]{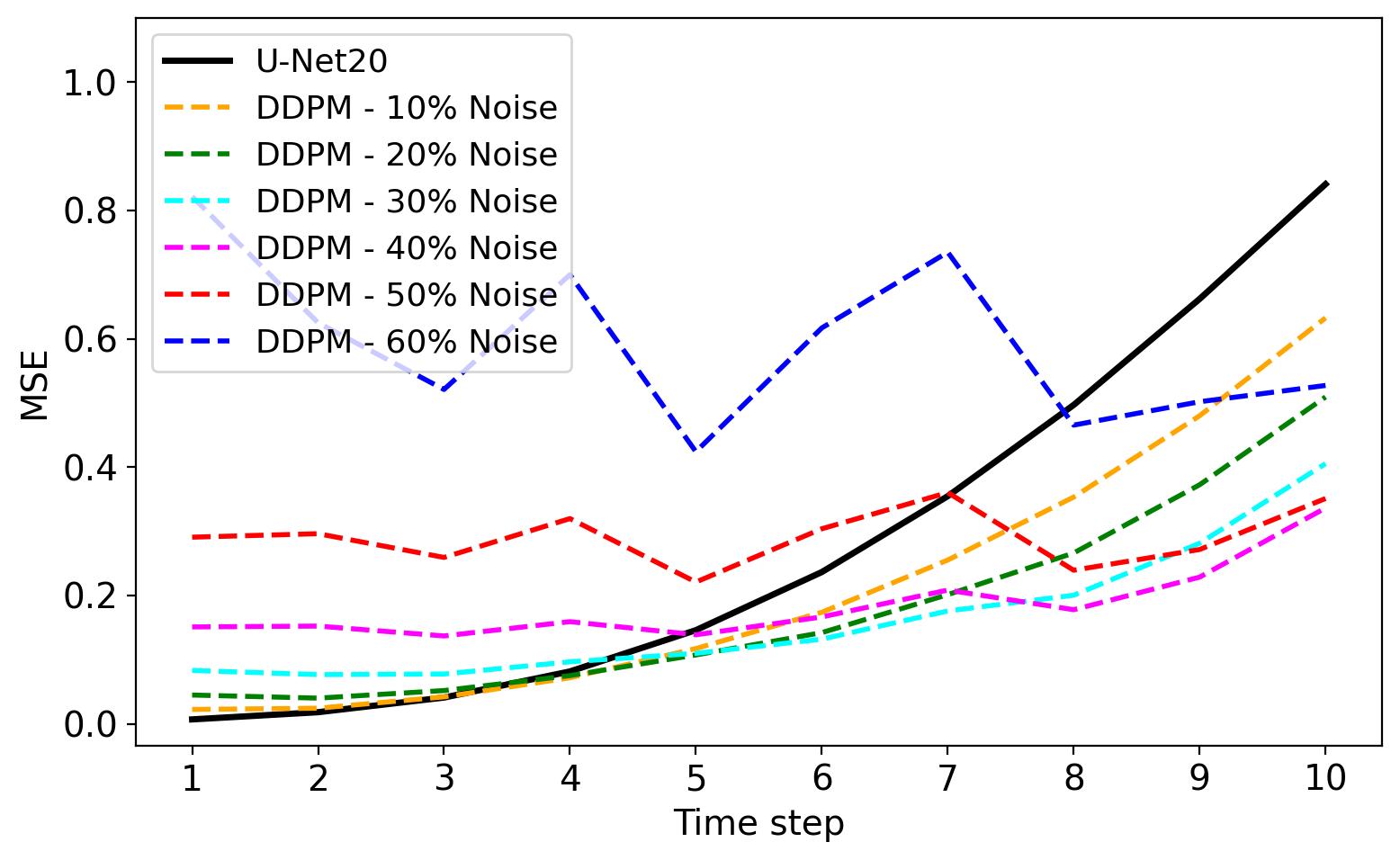}
        \label{fig:u-unet20}
    }

    \caption{Impact of different noise injection levels on mean squared error, MSE, for different models compared to CityFFD results for longitudinal velocity component: (a) CAE, (b) U-Net, (c) U-Net20}
    \label{Figureparametric}
\end{figure}

In the next step, the generated flow fields from each model are used as inputs to the DDPM to minimize spatial discrepancies and improve the flow field's fidelity.
In addition, to better understand the effects of DDPM implementation on the reconstructed flow fields of the DL models, it is important to analyze the effects of varying noise injection levels on the prediction accuracy of the CAE-LSTM and U-Net models.
As a result, the impact of noise levels of 10\%, 20\%, 30\%, 40\%, 50\%, and 60\% on the performance of predictive models is studied. Note that, for example, a noise level of 20\% corresponds to initiating the reverse process at step 200 out of the 1000 total steps in the noise addition process.
Figure \ref{Figureparametric} shows the mean squared error of the velocity component of each model after refinement by DDPM for each step of the predictions.

It should be noted that, in the CFD simulations of urban microclimate studies, reported wind speed errors typically range from 0.3 m/s to 1.5 m/s \cite{YANG2023110334}. As a result, in this study, to establish a consistent standard, a wind speed error of 1 m/s is utilized as a baseline for evaluating the performance of different DL models and assessing the effectiveness of DDPM implementation in reducing error propagation.
Consequently, the reported results in each figure are confined to the prediction step where the MSE of the velocity component is below this threshold.

As illustrated in Figure \ref{fig:u-cae}, DDPM has improved the prediction accuracy of longitudinal velocity over sequential prediction of the CAE model. However, because of the generative characteristic of the CAE's decoder, the improvement is not as significant as compared to the U-Net models (see Figures \ref{fig:u-unet} and \ref{fig:u-unet20}). Both DDPM and decoder function as a generative model and consistently guide the outputs back to the distribution of the training dataset. 

As shown in Figures \ref{fig:u-unet} and \ref{fig:u-unet20}, the implementation of the DDPM on both U-Net models has significantly improved their prediction accuracy in the longitudinal velocity component. 
In addition, errors accumulated at each step of the sequential prediction of U-Net models have the same effects as the incremental noise adding to the high-fidelity flow field. Therefore, reversing the diffusion process with deeper noise levels decreases the propagated error, significantly enhances the prediction accuracy, and reduces spatial discrepancies.
DDPM effectively models the turbulent flow by incrementally adding noise until a fully noisy image is achieved. Reversing this diffusion process from a deep noise level, especially close to the final step, creates an authentic representation of the fully developed turbulent flow \cite{lienen2024}. Therefore, as demonstrated in Figure \ref{Figureparametric}, a noise level of 60\% results in the generation of flow fields that are completely unrelated and less realistic compared to the actual instantaneous flow and exhibit higher errors compared to the reconstructed flow field of the DL models. Therefore, A balance between the amount of accumulated error and the noise injection level is required to enhance the effectiveness of DDPM in sequential time step prediction. 

Given the state-of-the-art accuracy of the U-Net model in one-step prediction and the substantial impact of DDPM on its performance, the U-Net model is selected as the foundation for visualizing velocity fields and evaluating the effectiveness of DDPM in the following sections.

\subsubsection{Optimal Noise Level for Error Reduction}

\begin {figure}[h!]
	\centering{\includegraphics[scale=0.45]{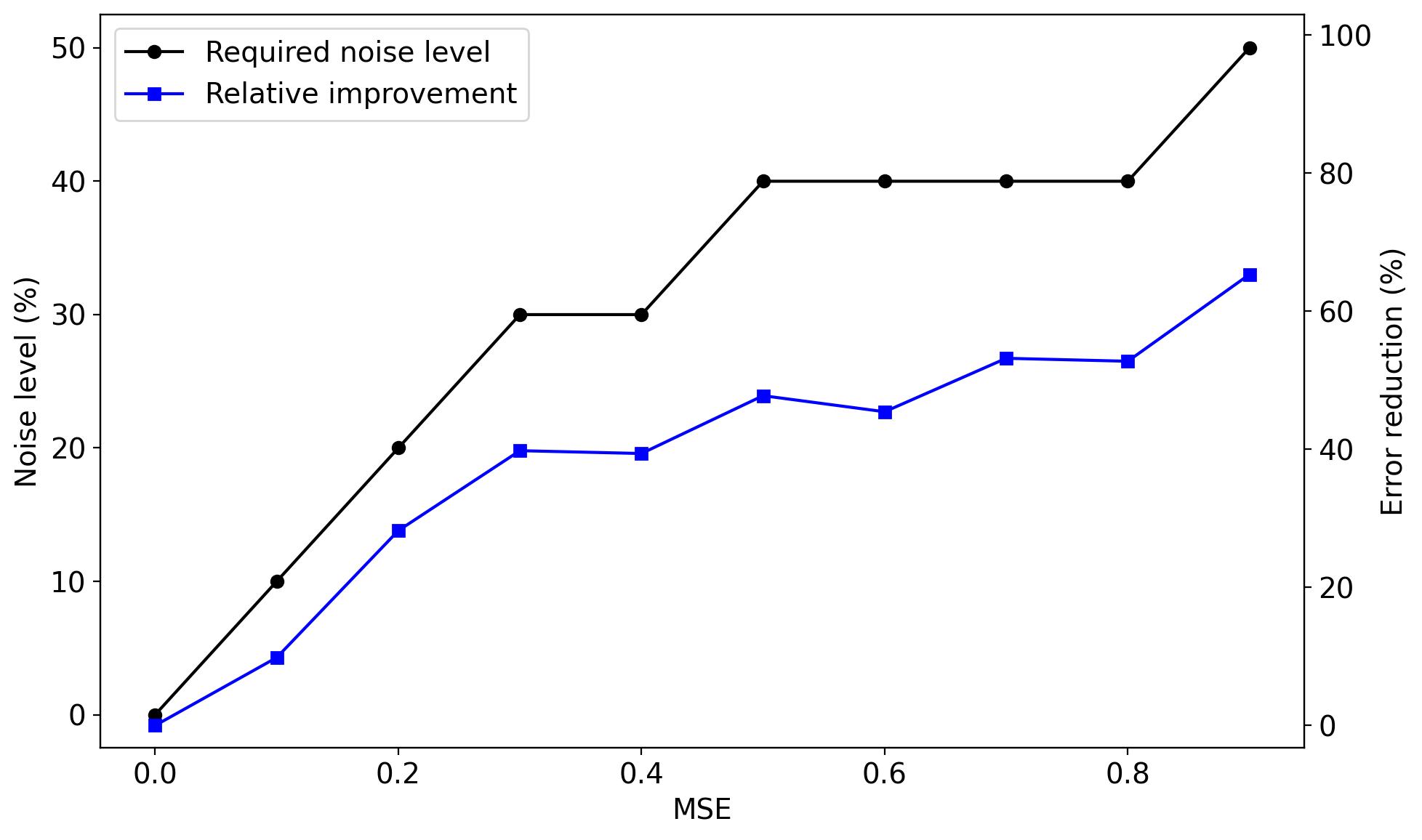}}
	\caption{Required noise level (\%) injection of reverse DDPM and relative error reduction (\%) as functions of MSE for the U-Net model's predictions. }
	\label{Figureoptimal_noise}
	\end{figure}

Figure \ref{Figureoptimal_noise} presents the required noise level for initiating the reverse process of DDPM as a function of MSE, alongside the relative error reduction achieved at each step in the U-Net model's predictions. 

The results indicate that the optimal starting noise level in DDPM is directly influenced by the accumulated error in the reconstructed flow field.
As the MSE increases and there are greater discrepancies in the predicted flow field, deeper noise levels are required to initiate the reverse DDPM process and effectively lower the error accumulation of the DL model. In addition, the relative error reduction consistently improves with increasing noise levels, achieving up to a 65\% reduction. This highlights the effectiveness of DDPM in addressing spatial discrepancies and minimizing error accumulation in sequential time-step predictions of DL models.
Consequently, a 50\% noise level injection is selected for further analysis of the longitudinal velocity component at time step 6417, where the U-Net model reaches the error threshold.

\begin{figure}[h!]
    \centering

    \textbf{Time step = 6417}\\[2mm]

    \raisebox{-.5\height}{\rotatebox{90}{\textbf{Ground truth}}}
    \begin{minipage}[c]{0.4\textwidth}
        \centering
        \includegraphics[width=\linewidth]{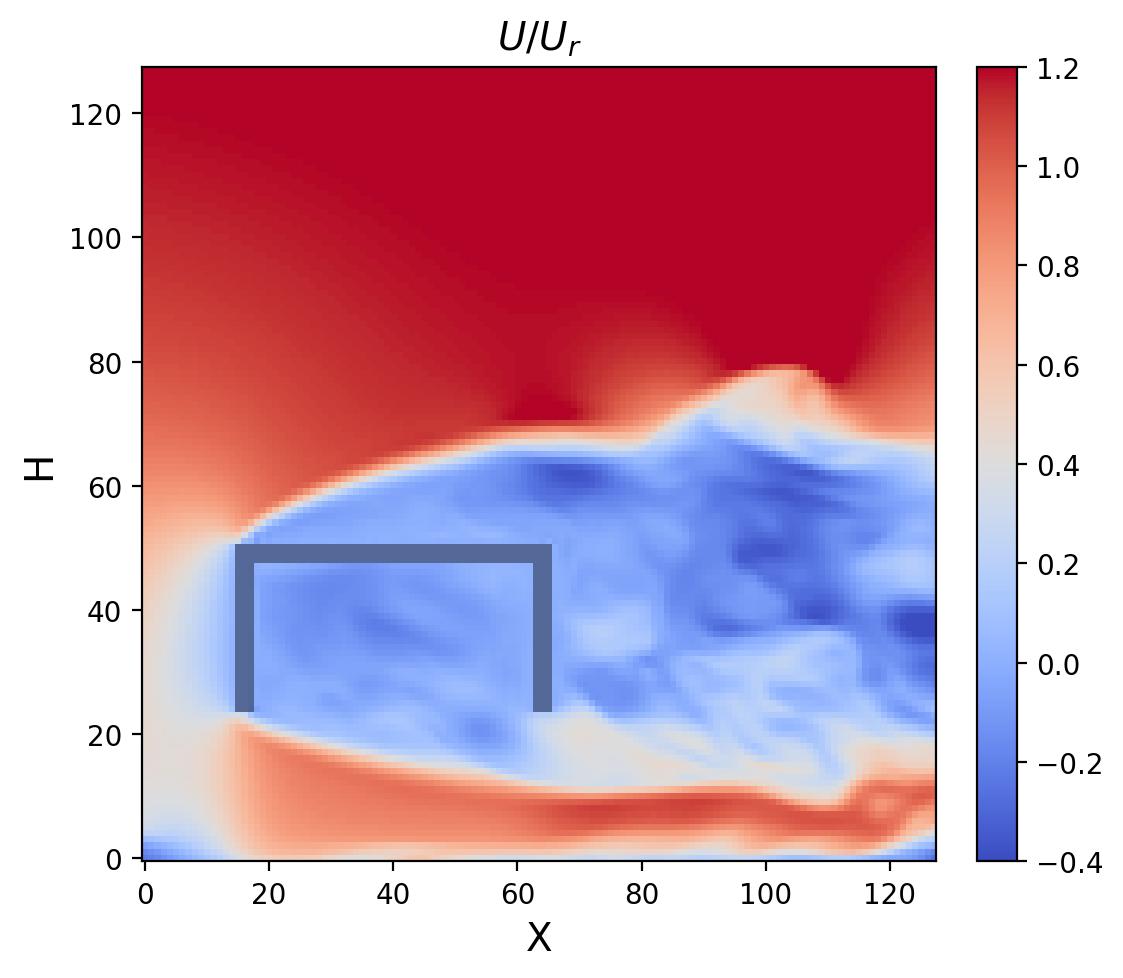}
    \end{minipage}

    \begin{minipage}[b]{0.4\textwidth}
        \centering
        \textbf{U-Net(MSE = 0.927)}
    \end{minipage}
    \hspace{5mm}
    \begin{minipage}[b]{0.4\textwidth}
        \centering
        \textbf{U-Net-DDPM(MSE = 0.321)}
    \end{minipage}

    \raisebox{-.5\height}{\rotatebox{90}{\textbf{Prediction}}}
    \begin{minipage}[c]{0.4\textwidth}
        \centering
        \includegraphics[width=\linewidth]{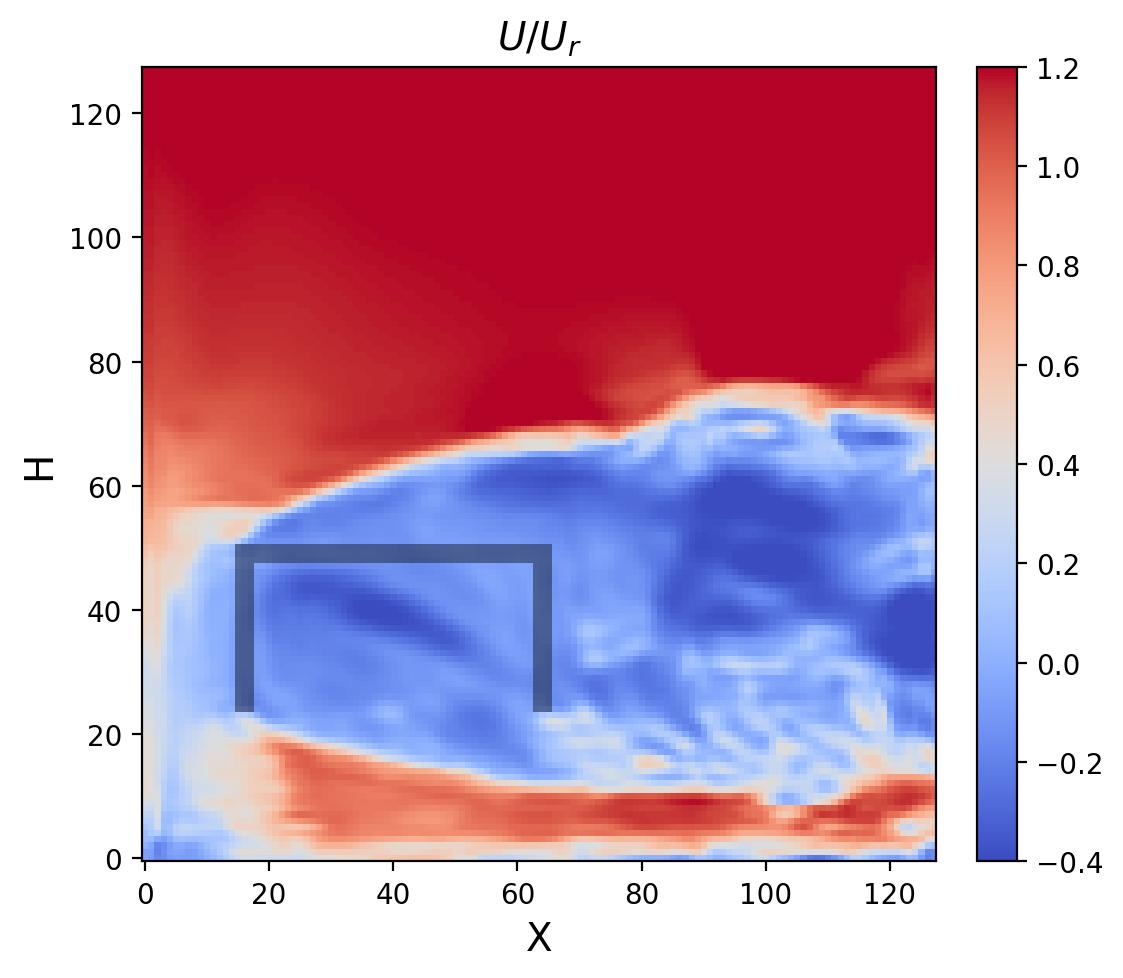}
    \end{minipage}
    \hspace{5mm}
    \begin{minipage}[c]{0.4\textwidth}
        \centering
        \includegraphics[width=\linewidth]{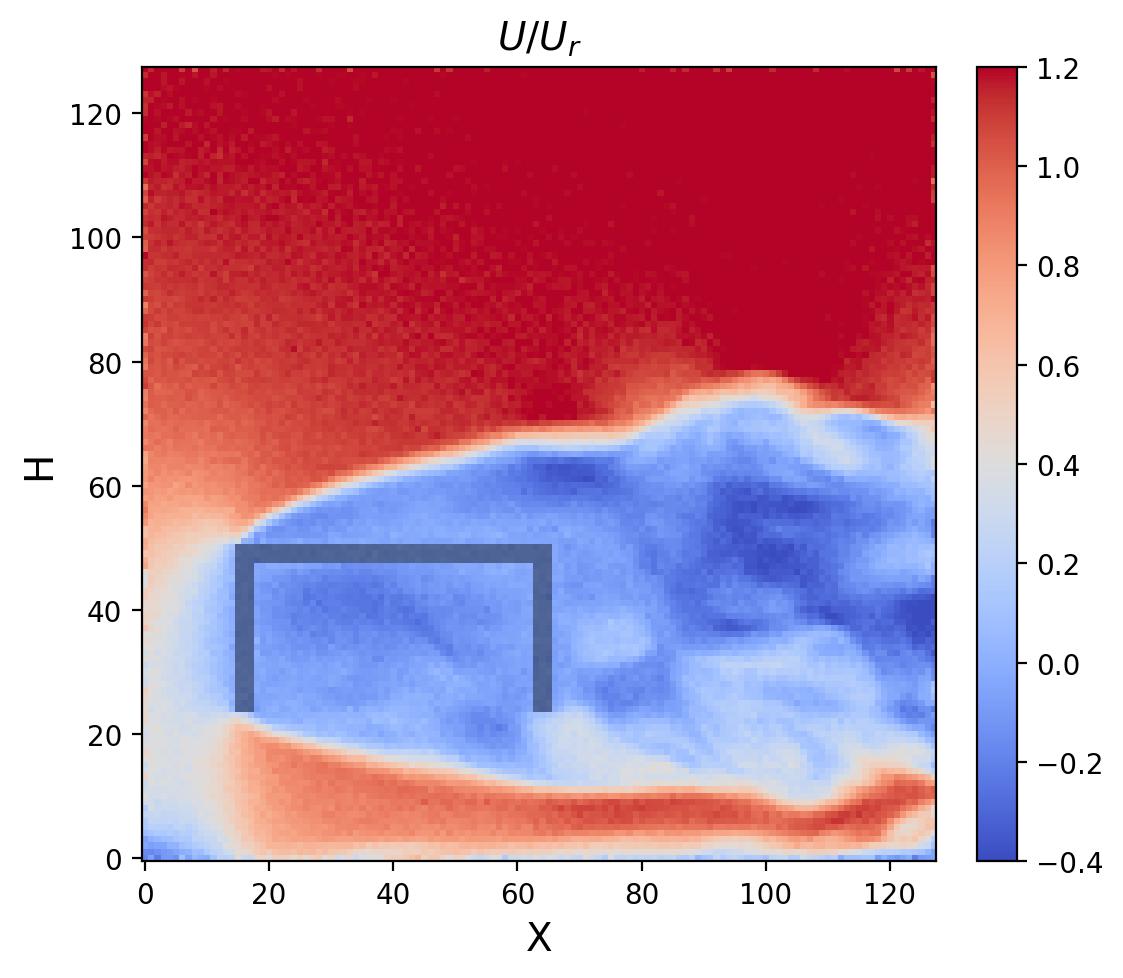}
    \end{minipage}

    \raisebox{-.5\height}{\rotatebox{90}{\textbf{Absolute error}}}
    \begin{minipage}[c]{0.4\textwidth}
        \centering
        \includegraphics[width=\linewidth]{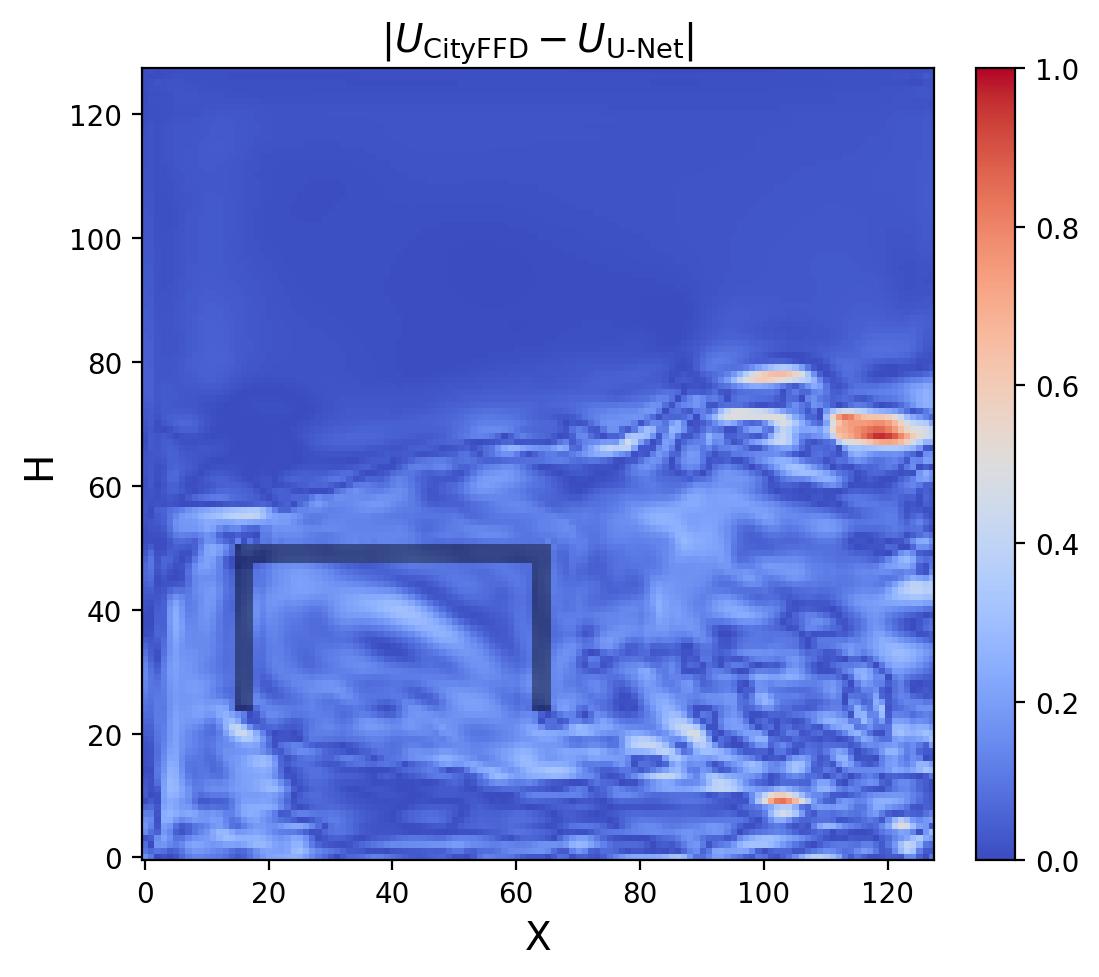}
    \end{minipage}
    \hspace{5mm}
    \begin{minipage}[c]{0.4\textwidth}
        \centering
        \includegraphics[width=\linewidth]{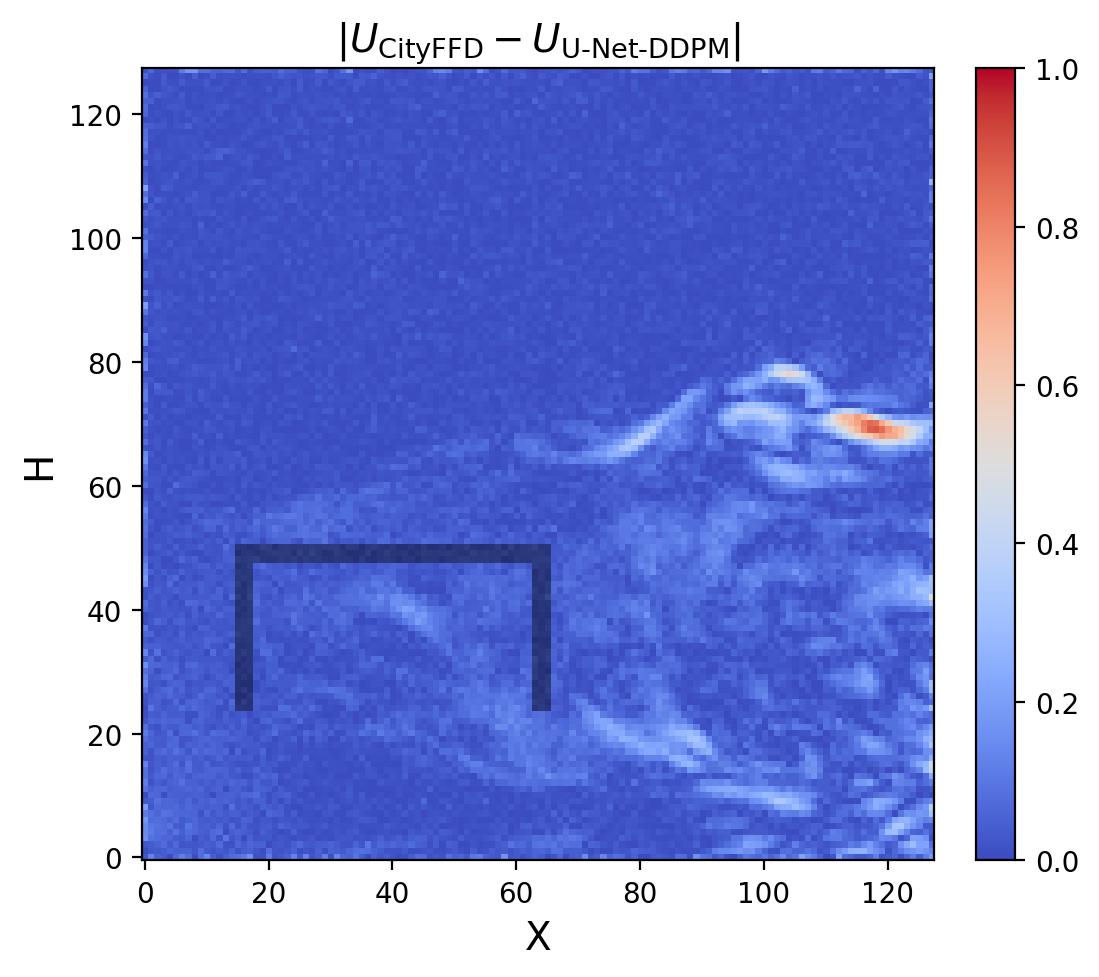}
    \end{minipage}

    \caption{The contours of the normalized streamwise velocity, reconstructed by the U-Net, U-Net-DDPM, and the CityFFD data at time step 6417, along with the contour of absolute deviation of the dimensionless longitudinal velocity.}
    \label{Figureforecast_u}
\end{figure}

Figure \ref{Figureforecast_u} presents the contour visualization of the dimensionless longitudinal velocity, along with the absolute error contours of the longitudinal velocity component reconstructed by U-Net and U-Net-DDPM compared to the CityFFD data at time step 6417.
\clearpage

The longitudinal velocity component reconstructed by U-Net deviates from the CityFFD data due to the error accumulation inherent in the model's sequential predictions, resulting in noticeable discrepancies from the ground truth.
However, the DDPM significantly enhances the accuracy of the reconstructed flow field.
As a result of high fluctuations in the streamwise velocity component, the DDPM is trained over a wide range of velocity distributions, which enables the model to be more effective in reducing the discrepancies attributed to the U-Net results. As shown in Figure \ref{Figureforecast_u}, DDPM enhances the model's accuracy, especially in the regions around the building block.
However, slight deviations compared to ground truth are observed in the wake region, where the flow is fully turbulent. 

\subsubsection{Error Distribution}

\begin {figure}[h!]
	\centering{\includegraphics[scale=0.6]{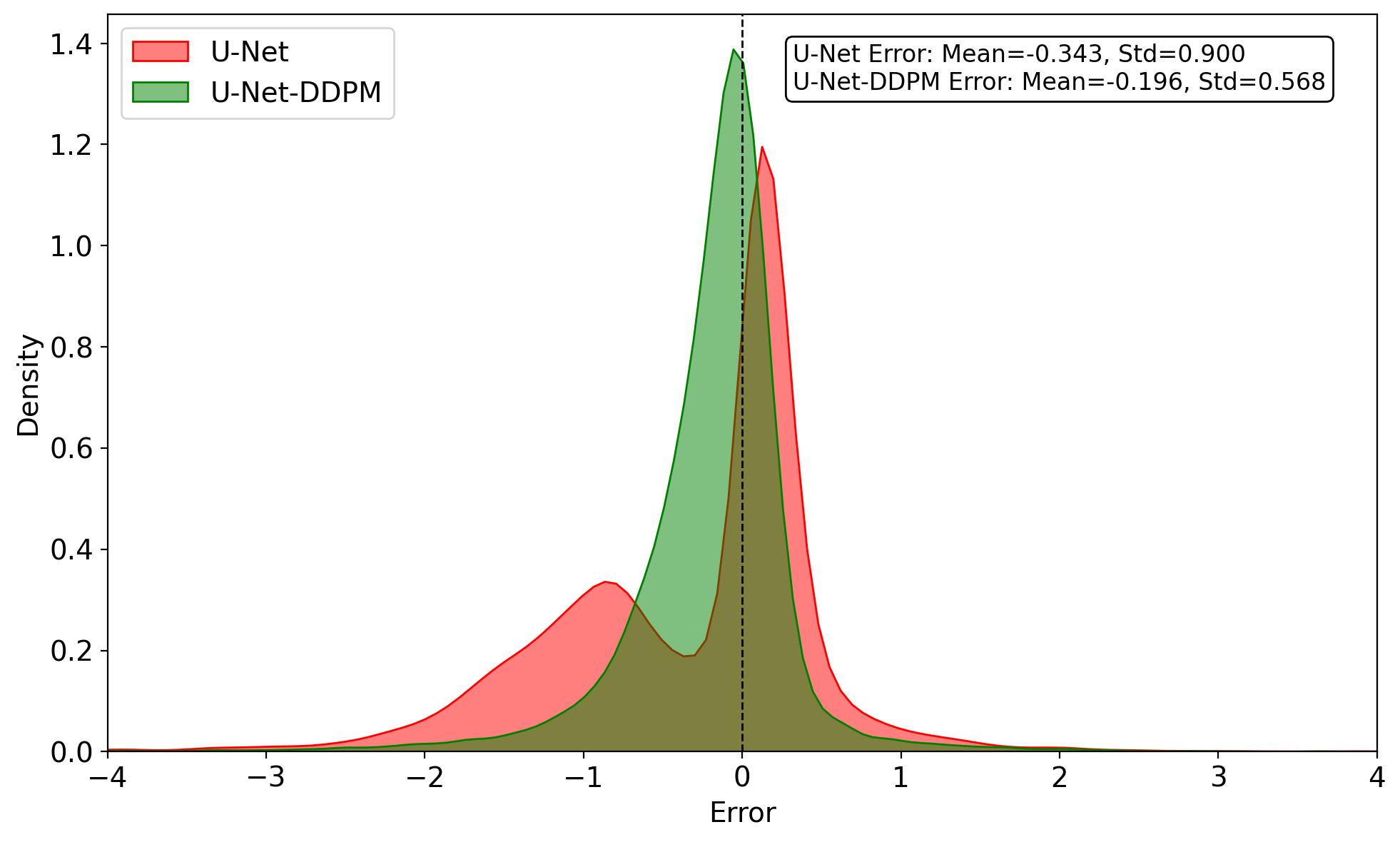}}
	\caption{Error distribution of U-Net and U-Net-DDPM at time step 6417.}
	\label{Figureerror-distribution}
	\end{figure}

To better study the performance of DDPM, Figure \ref{Figureerror-distribution} shows the error distribution of the reconstructed flow fields of the U-Net Model and the refined flow field of U-Net-DDPM at time step 6417. A red curve represents the error density of the U-Net model with a mean error of -0.343 and a standard deviation of 0.900. In contrast, the green curve, representing the refind flow fields of U-Net-DDPM, exhibits a narrower distribution with a mean error of -0.196 and a standard deviation of 0.568. 
Implementation of DDPM significantly improves the reconstructed flow fields of the U-Net model by effectively reducing high-magnitude errors and aligning the predictions closer to the ground truth. Furthermore, DDPM minimizes variation in prediction errors due to its narrower and more centered error distribution. 

\subsubsection{Energy Spectrum  }

\begin {figure}[h!]
	\centering{\includegraphics[scale=0.6]{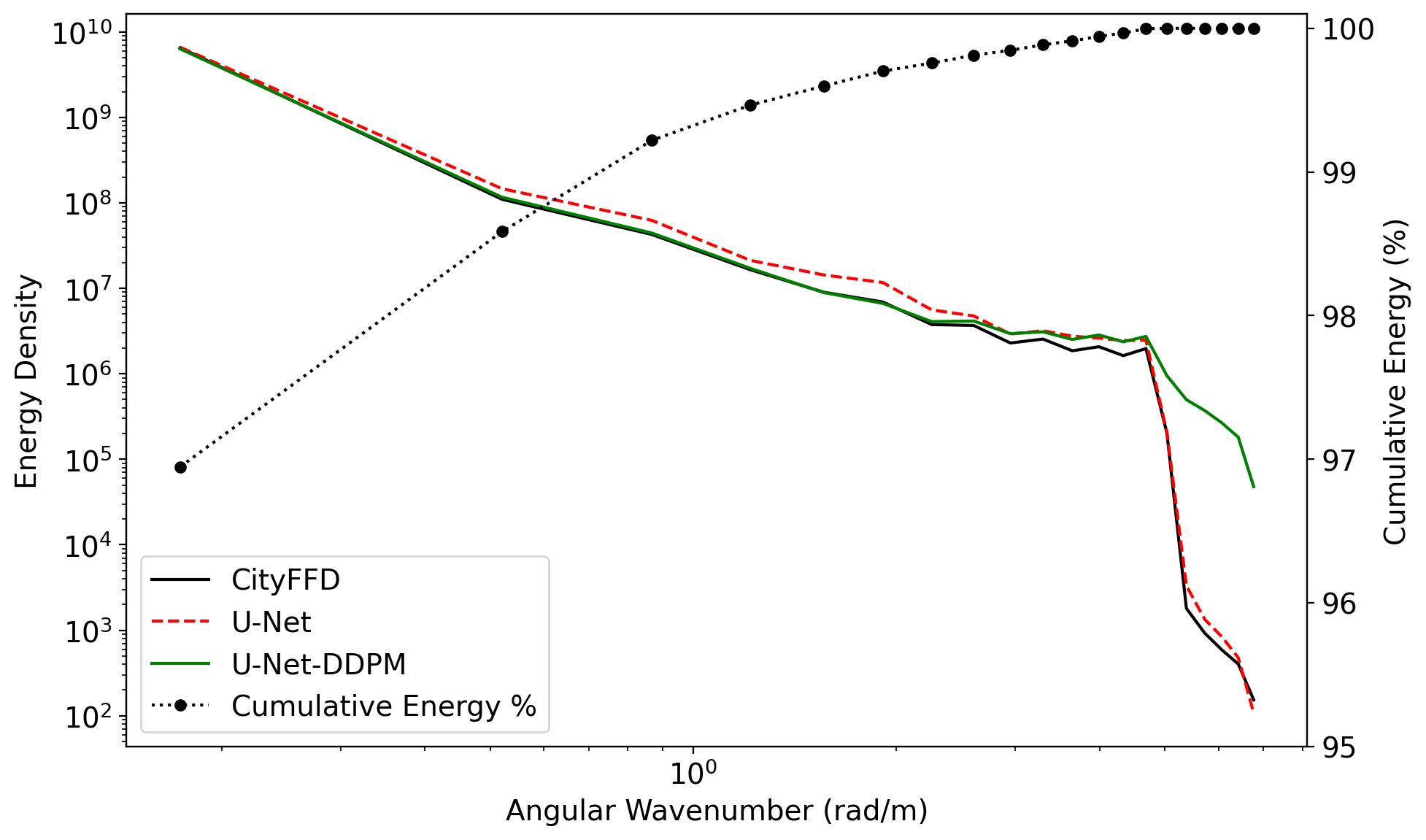}}
	\caption{Power Spectral Density and Cumulative Energy Distribution of CityFFD, U-Net, and U-Net-DDPM at time step 6417.}
	\label{FigurePSD}
	\end{figure}

To further analyze the performance of DDPM, Figure \ref{FigurePSD} depicts the power spectral density (PSD)\cite{kolmogorov1941local} and cumulative energy distribution of the ground truth data at time step 6417. It is shown that DDPM implementation on the reconstructed flow fields improves the model's prediction significantly. In the high-energy density regions (lower wave numbers), U-Net-DDPM aligns closely with the CityFFD data, effectively capturing the dominant flow structures that contribute the most to the total energy distribution. At higher wave numbers, U-Net-DDPM results deviate from the ground truth due to the small error remaining in the domain after the denoising process. The cumulative energy curve, however, demonstrates that these deviations have a relatively small impact on the system's total energy. 

\subsection{Computational efficiency comparison}

The computational efficiency of the numerical solver is compared with DL methods employed in this study, including CAE, U-Nets, and DDPM. The evaluation is conducted on an NVIDIA GTX TITAN V graphics card, with computation time measured in seconds for processing a single forward step in the flow field. It is important to note that the accuracy of this comparison may be inaccurate, as the DL models are trained using a 2D dataset, while the numerical solver, despite being optimized for GPU usage, processes 3D outputs. Table \ref{table.comp-time} indicates that the DL approaches provide significant computational advantages over the traditional numerical solvers. In addition, while implementing the DDPM on the result of the predictive models increases the computation cost, it still provides around 3 times speedup compared to the numerical simulation, while maintaining acceptable accuracy for sequential time step predictions.

\begin{table}[h!]
\centering
\caption{Computational efficiency performance.}
\label{table.comp-time}
\begin{tabular*}{\textwidth}{@{\extracolsep{\fill}}lccccc}
\hline
\multicolumn{1}{c}{} & \multicolumn{5}{c}{Methodology} \\
\cline{2-6}
Characteristic & CityFFD & CAE-LSTM & U-Net & U-Net20 & DDPM \\
\hline
GPU time (s) & 0.485 & 0.003 & 0.001 & 0.008 & 0.175 \\
\hline
\end{tabular*}
\end{table}

\section{Conclusions}
\label{sec.conclusion}

This study has introduced an application of the DDPM as a post-processing technique for DL models, specifically CAE-LSTM and U-Net, to reduce the spatial discrepancies of airflow prediction over a cubic single building with openings at the windward and leeward sides under fully turbulent conditions. It is demonstrated that DDPM can reconstruct high-fidelity turbulent flow fields from low-fidelity images generated by DL models in sequential time-step prediction applications. Notably, it achieves up to a 65\% improvement in accuracy compared to the DL models alone, while maintaining acceptable accuracy relative to CFD data and providing 3 times acceleration compared to the numerical solver. By learning the underlying probability distribution of the data, DDPM effectively captures complex turbulent features and accurately models the statistical and physical properties of turbulent flows, including their inherent randomness. Additionally, the deterministic nature of DDIM enables faster sampling and allows the reconstruction of deterministic samples at any step of the reverse denoising process. 

The interaction between the DL model type and the DDPM’s performance also emerged as an important factor in enhancing the fidelity of the reconstructed flow fields. In particular, the generative behavior of the CAE's decoder, which aims to constrain outputs within the trained data distribution, limits the DDPM’s potential for further enhancement. However, the DDPM implemented on U-Net significantly improved the prediction accuracy for both velocity components. This illustrates the DDPM’s utility in denoising cumulative errors throughout the sequential prediction process and aligning the outputs closer to the CFD data distributions.

Adding deeper noise levels to the DDPM can also enhance the accuracy and reduce spatial discrepancies, especially as models iterate through more steps and accumulate more errors. By starting the reverse diffusion from deeper noise levels, the fully sorted predictions are refined, and the overall reconstruction accuracy is enhanced. However, a balance must be maintained, as initiating the reverse denoising process from a deeper level can lead to less realistic flow field reconstructions.

The error distribution analysis demonstrates that DDPM effectively reduces high-magnitude errors, resulting in a narrower and more centered error distribution. Additionally, the PSD analysis confirms that DDPM preserves dominant flow structures, highlighting its capability to improve the accuracy of generated flow fields while maintaining essential flow characteristics.

To refine the predictions of different predictive models, a 2D DDPM is used, however, to extend the proposed method for real-world applications, a 3D DDPM needs to be developed. Also, currently, DDPM refines flow fields independently from the DL model. Integrating DDPM into the predictive pipeline of DL models could improve accuracy over extended time sequences. These potential ways to resolve the limitations of DDPM to enhance the predictive model’s performance will be the main focus of future works.

\section*{CRediT authorship contribution statement}

\textbf{Sepehrdad Tahmasebi}, \textbf{Geng Tian}, and \textbf{Shaoxiang Qin}:
Writing – original draft, Writing - Review \& editing, Formal analysis, Visualization, Validation, Methodology, Investigation, Conceptualization. \textbf{Ahmed Marey} and \textbf{Saeed Rayegan}: Review \& editing, Methodology, Formal analysis, Investigation. \textbf {Liangzhu (Leon) Wang}: Review \& editing, Conceptualization, Supervision, Resources, Funding acquisition.

\section*{Declaration of competing interest}

The authors declare that they have no financial or non-financial conflicts of interest.

\section*{Data availability}

Additional data supporting the findings of this study are available from the corresponding author upon reasonable request.

\section*{Acknowledgements}
\addcontentsline{toc}{section}{Acknowledgements}
The research is made possible by the general support from the Consortium for Research and Innovation in Aerospace in Québec (CRIAQ) through the project "Multi-scale Aerodynamic Modeling of Helicopters/UAV in Urban Environments"[\#051811], the Natural Sciences and Engineering Research Council (NSERC) of Canada through the Discovery Grants Program [\#RGPIN-2024-06297], and the SEED project “Creating Electrified and Decar-bonized Healthy Urban Microclimate around Building Clusters through Climate-Resilient Solu-tions” under the Canada First Research Excellence Fund (Volt-Age).

\appendix
\numberwithin{equation}{section}
\renewcommand{\thesection}{\Alph{section}}
\section{DL Models Architecture}
\label{sec.appendixA}
\setcounter{figure}{0}

\renewcommand{\thefigure}{A.\arabic{figure}}

\setcounter{table}{0}

\renewcommand{\thetable}{A.\arabic{table}}
\subsection*{CAE-LSTM model}

In this study, downsampling blocks (DS blocks) are designed with convolutional layers featuring strides greater than one, as opposed to traditional pooling layers. This approach leverages convolutional strides for spatial downsampling, a method shown to preserve spatial information more effectively \cite{zhu2021stacked}. Specifically, each DS block applies a convolution layer with a stride of (2, 2) and a kernel size of (3, 3), facilitating a more controlled reduction in spatial dimensions based on the desired downsampling factor. This design allows for an adaptive approach to downsampling, where the stride size directly correlates with the scale of dimensionality reduction.
In the decoder portion, the network reconstructs the input data by leveraging upsampling blocks (UP blocks), each incorporating a transpose convolution layer. These layers mirror the downsampling design, with a stride of (2, 2) and a kernel size of (3, 3), gradually reconstructing the data from the latent space back to its initial shape.
Further enhancing the model, each DS and UP block is succeeded by a residual convolution block (RES block). Each RES block comprises two convolutional layers with a stride of (1, 1) and a kernel size of (3, 3). The inclusion of residual connections within these blocks improves the model’s performance by mitigating potential vanishing gradient issues and allowing for deeper, more robust feature learning. These residual connections thus contribute to stable training and improved feature representation throughout the autoencoder architecture.

	\begin {figure}[h!]
	\centering{\includegraphics[scale=0.5]{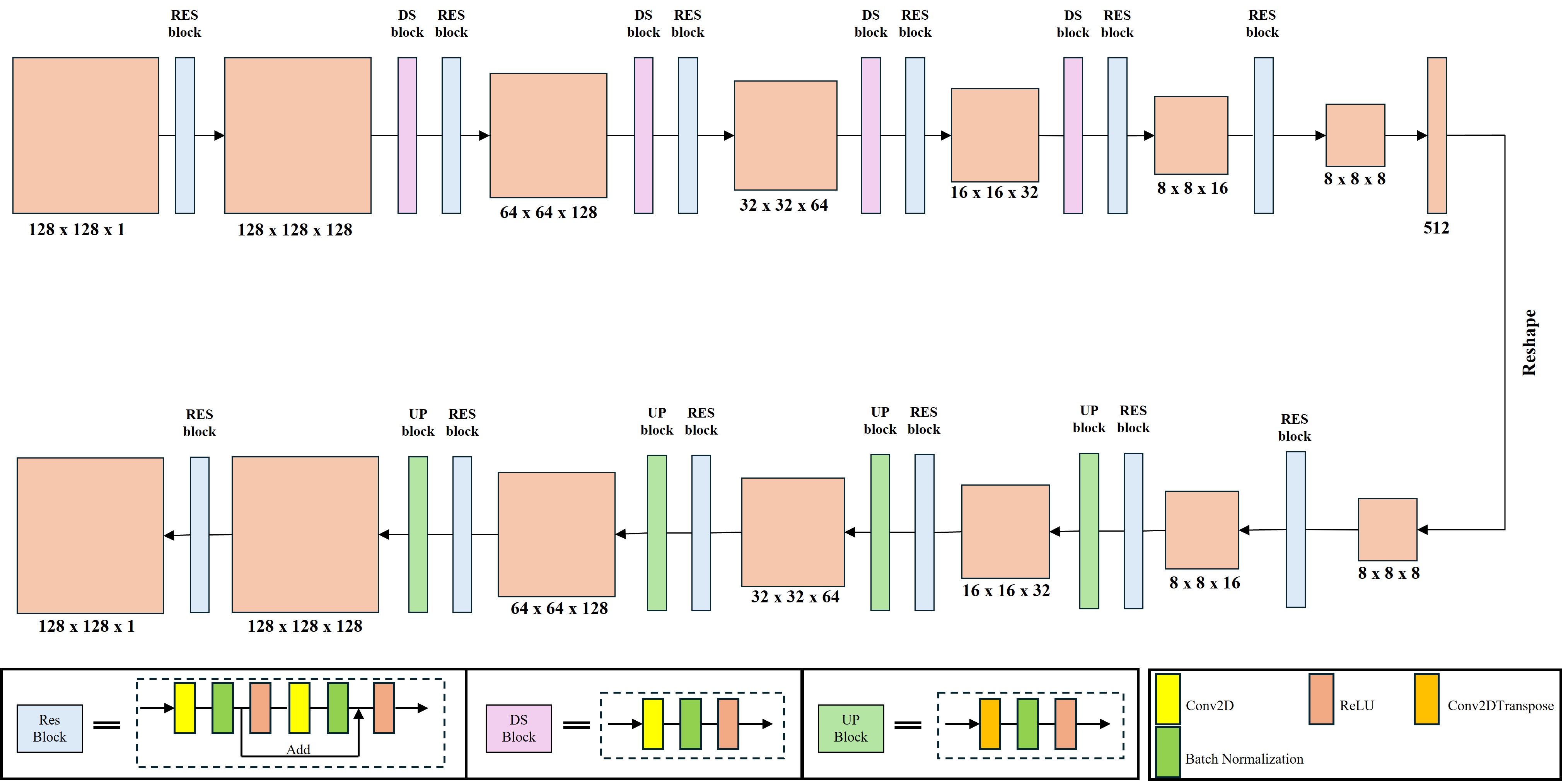}}
	\caption{An overview of the structure of the CAE used in the present study. }
	\label{Figureautoencoder}
	\end{figure}

\begin {figure}[h!]
	\centering{\includegraphics[scale=0.7]{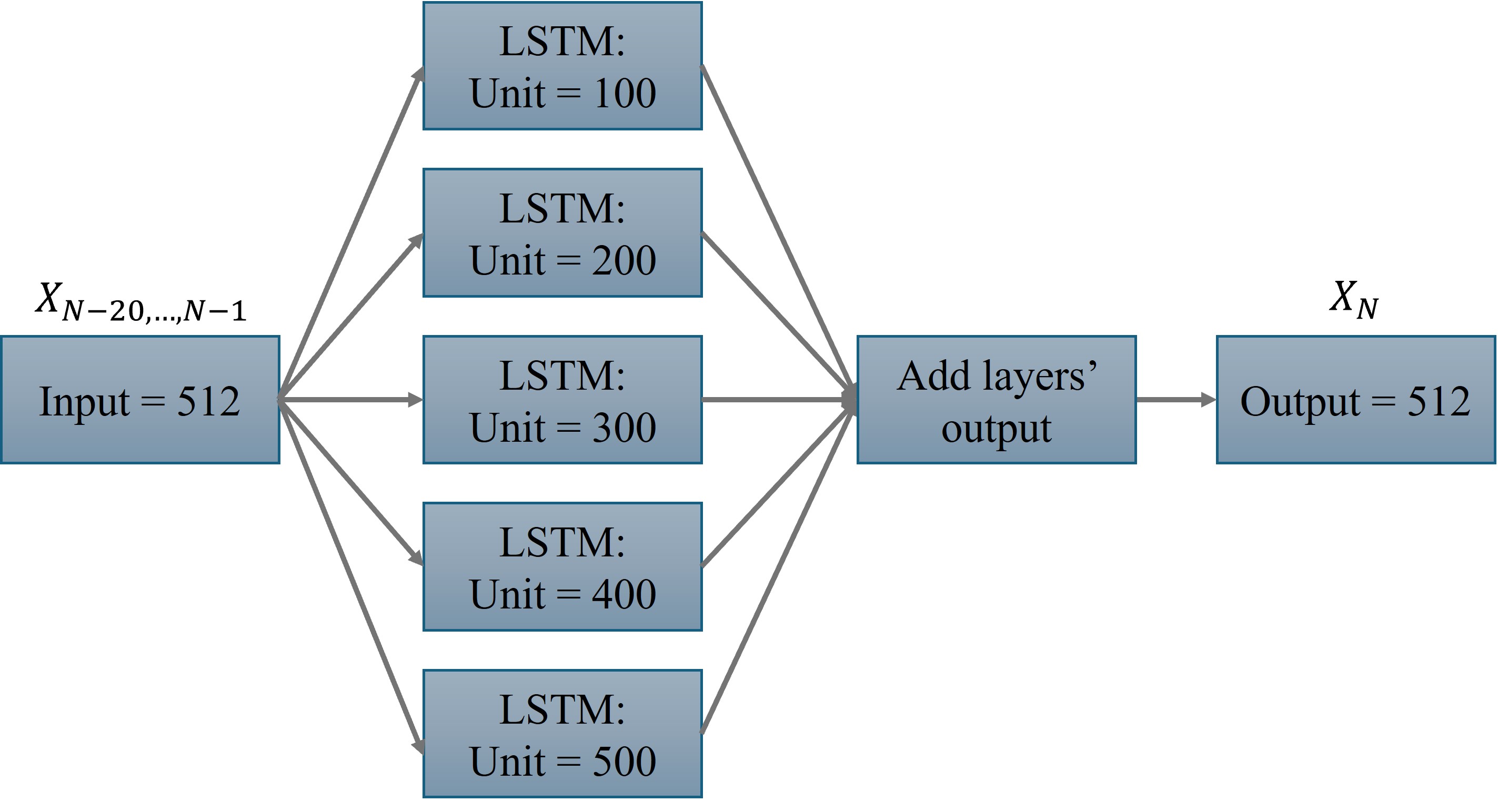}}
	\caption{Parallel LSTM architecture utilized in the present study.}
	\label{Figureparallel LSTM}
	\end{figure}

Regarding the modeling of the temporal development of the latent space obtained from the CAE, the structure of parallel LSTM used in the present study is depicted in Figure \ref{Figureparallel LSTM}. Various configurations are tested for accuracy and computation time, including different numbers of layers, hidden neurons, and sequence lengths. Specifically, sequence lengths of 5, 10, 15, and 20 are evaluated. The results indicated that a sequence length of 20 provided the best performance, suggesting that the model’s output effectively captures information from the previous 20 time steps (detailed results not shown here). Moreover, changes in the number of layers and hidden neurons had minimal effect on performance. As a result, the parallel LSTM architecture in the present study utilizes five parallel layers with units of 100, 200, 300, 400, and 500, respectively. 

\subsection*{U-Net models}

The down-sampling phase employs a max-pooling layer with a stride of 2, followed by two convolutional layers with a (1, 1) stride and a (3, 3) kernel size. Conversely, upsampling is achieved through a transposed convolutional layer with a stride of (2, 2) and a (3, 3) kernel size, succeeded by two convolutional layers with a (1, 1) stride and a (3, 3) kernel size. These model configurations aim to enhance predictive accuracy in complex flow field reconstructions.

	\begin {figure}[h!]
	\centering{\includegraphics[scale=0.61]{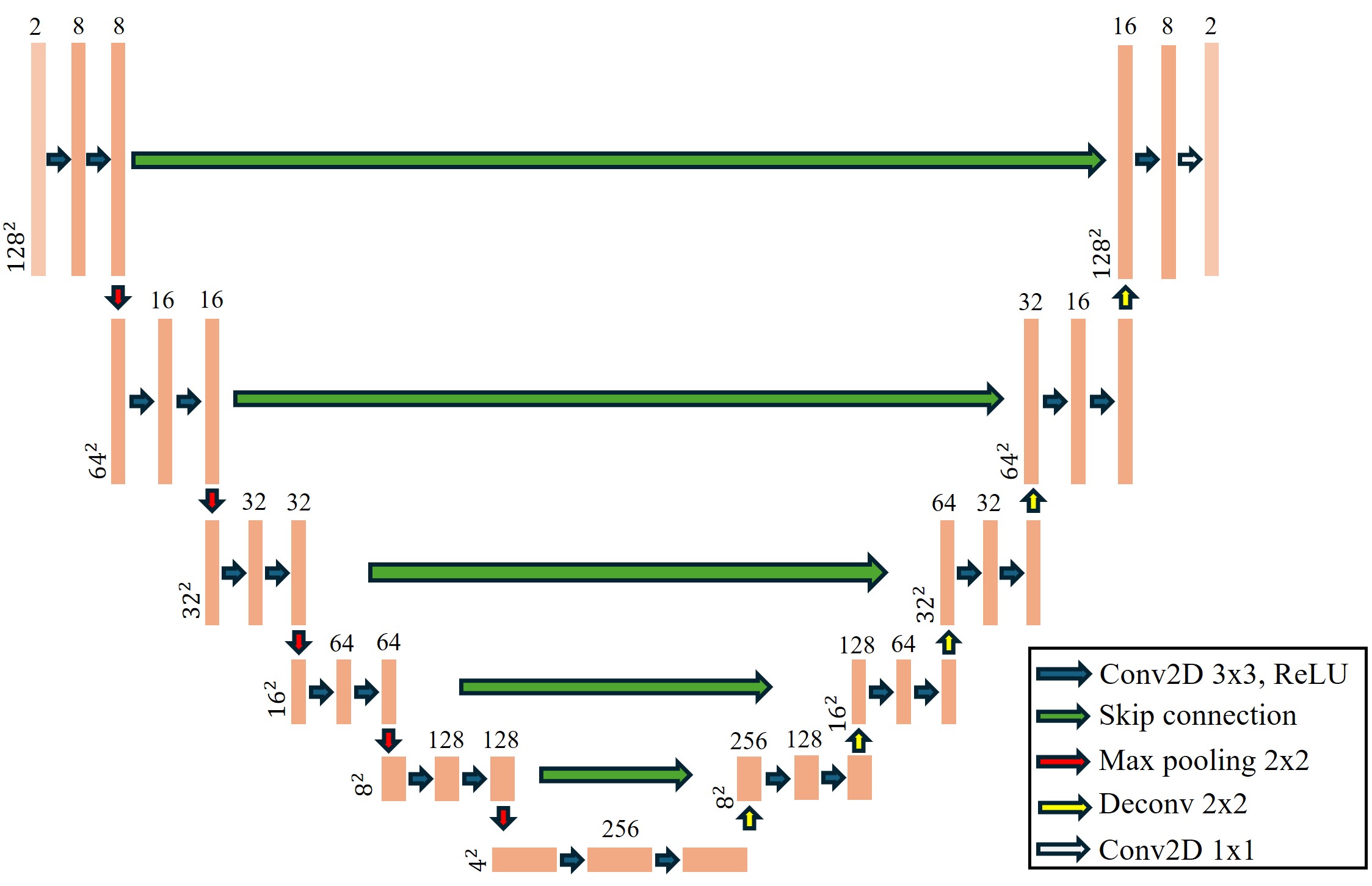}}
	\caption{An overview of the U-Net architecture used in the present study. Each convolutional layer includes activation and BatchNormalization layers.}
	\label{Figureunet}
	\end{figure} 

\begin{table}[h!]
\centering
\caption{Number of channels in each layer of the U-Net20 model.}
\begin{tabular*}{\textwidth}{@{\extracolsep{\fill}}cccc}
\hline

Model's path & Layer & Input Channel & Output Channel \\
\hline
\multirow{6}{*}{Encoder} 
& 1 & 40 & 64 \\

& 2 & 64 & 128 \\

& 3 & 128 & 256 \\

& 4 & 256 & 512 \\

& 5 & 512 & 1024 \\

& 6 & 1024 & 2048 \\
\hline
\multirow{6}{*}{Decoder} 
& 1 & 2048 & 1024 \\

& 2 & 1024 & 512 \\

& 3 & 512 & 256 \\

& 4 & 256 & 128 \\

& 5 & 128 & 64 \\

& 6 & 64 & 2 \\
\hline
\end{tabular*}
\label{tab:unet20}
\end{table}
\clearpage
\subsection*{DDPM}
	\begin {figure}[h!]
	\centering{\includegraphics[scale=0.5]{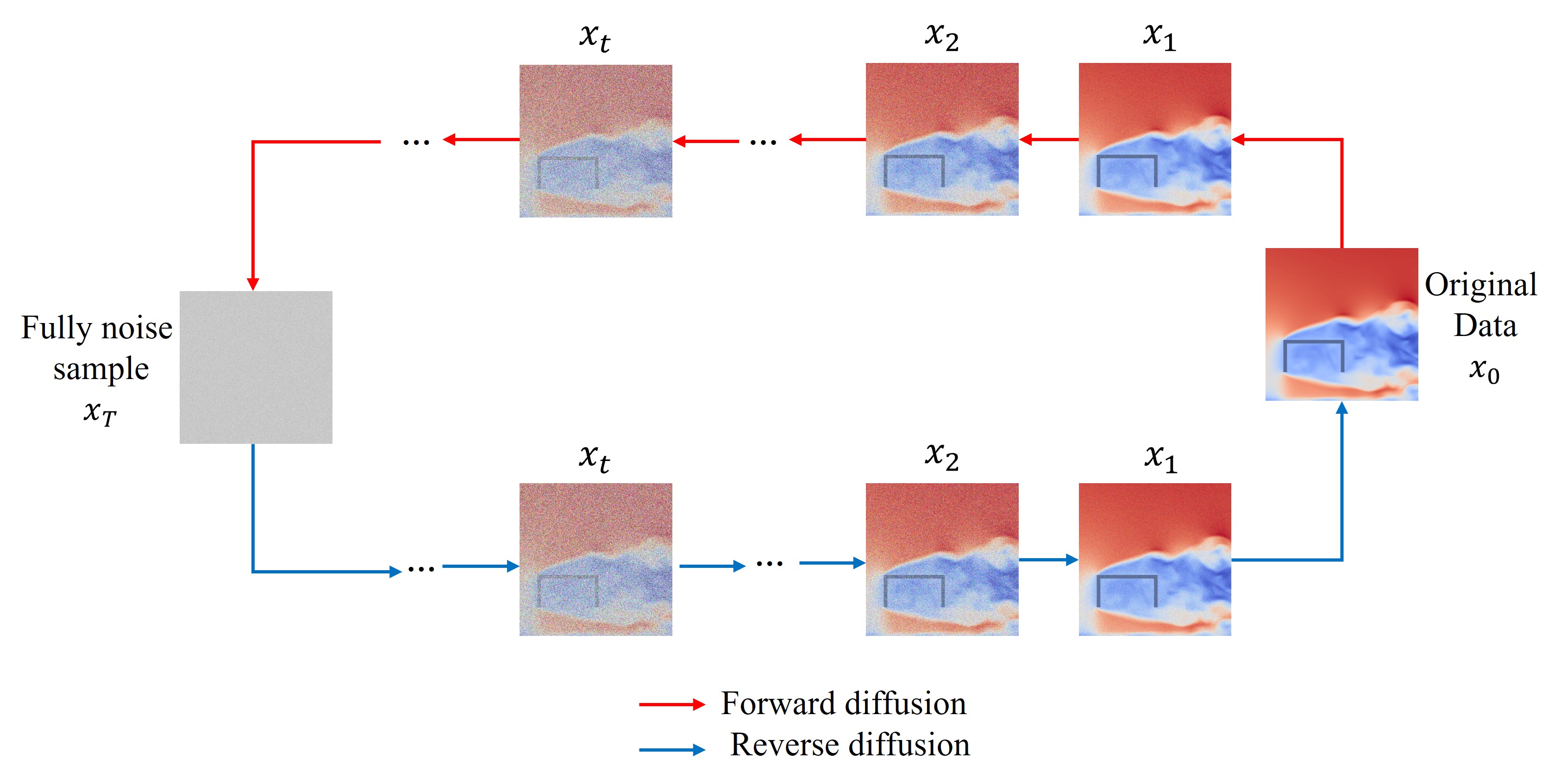}}
	\caption{An overview of a DDPM model: Red arrows represent the forward diffusion process by adding noise at each step, and blue arrows indicate reverse diffusion by removing the noise at each step to finally get an authentic sample. }
	\label{Figureddpm}
	\end{figure} 

The forward process involves progressively adding Gaussian noise to the clean image at each time step, transforming it into a fully noisy version, with the variance schedule $(\beta_t)$ controlling the noise added at each step. Let $x_0$ represent the clean image; the forward diffusion process over $T$ steps can be expressed as follows \cite{ho2020}:

\begin{equation}
\begin{aligned}
q(x_t \mid x_{t-1}) &:= \mathcal{N}(x_t; \sqrt{1 - \beta_t}x_{t-1}, \beta_t I), \quad
q(x_{1:T} \mid x_0) &:= \prod_{t=1}^T q(x_t \mid x_{t-1})
\end{aligned}
\label{eq:both_equations}
\end{equation}

Since the data is obtained based on a sample from a Gaussian distribution, the distribution of $x_t$ given $x_0$ can be defined as follows \cite{ho2020}:

\begin{equation}
\begin{aligned}
q(x_t \mid x_0) &:= \mathcal{N}(x_t; \sqrt{\bar{\alpha}_t} x_0, (1 - \bar{\alpha}_t)I), \\
\text{where } \alpha_t &:= 1 - \beta_t \quad \text{and} \quad \bar{\alpha}_t := \prod_{i=1}^t \alpha_i.
\label{ddpm-equ}
\end{aligned}
\end{equation}

During the backward diffusion process, the target is to create an authentic image sample $x_0$. The reverse process is defined as \cite{ho2020}:

\begin{equation}
p_\theta (x_{0:(T-1)} \mid x_T) := \prod_{t=1}^T p_\theta (x_{t-1} \mid x_t), \quad p_\theta (x_{t-1} \mid x_t) := \mathcal{N}(x_{t-1}; \mu(x_t, t), \sigma_t^2 I)
\end{equation}

where $\theta$ is the set of neural network parameters of the DDPM model, $\sigma_t^2$ is the variance, and $\mu_\theta$ is the mean that is parametrized by the neural network.
The main objective of the reverse process is to maximize the likelihood probability of $p_\theta(x_0)$. 
In variational inference, the combination of $q$ and $p$ forms the basis of VAEs \cite{kingma2013auto}, and the variational lower bound can be expressed as: 

\begin{equation}
L_{\text{VLB}} := L_0 + L_1 + \dots + L_{T-1} + L_T 
\label{eq.ariational lower bound}
\end{equation}

Where
\begin{equation}
\begin{aligned}
L_0 &:= -\log p_{\theta}(x_0 | x_1), \\ 
L_{t-1} &:= D_{KL}(q(x_{t-1} | x_t, x_0) \parallel p_{\theta}(x_{t-1} | x_t)), \\ 
L_T &:= D_{KL}(q(x_T | x_0) \parallel p(x_T)).
\end{aligned}
\label{eq:combined_equation}
\end{equation}

Note that aside from \(L_0\), all the terms in Eq.~(\ref{eq.ariational lower bound}) represent a KL divergence between two Gaussian distributions. In Eq.~(\ref{eq:combined_equation}), \(L_T\) corresponds to the forward process of the diffusion model, while \(L_{1:T-1}\) characterizes the reverse diffusion process. Additionally, to evaluate \(L_0\) for images, each pixel value, originally ranging from 0 to 255, is linearly scaled to \([-1, 1]\).
This scaling ensures that the inputs to the neural network's reverse process are consistent, facilitating effective training.
As a result, the simplified training objective of the diffusion model, which focuses on parametrizing the Gaussian probability density function while ignoring the constant variances \(\beta_t\), is given by \cite{ho2020}:

\begin{equation}
L := \mathbb{E}_{t, x_0, \epsilon} \left[ \left\| \epsilon_t - \epsilon_{\theta} \left( \sqrt{\bar{\alpha}_t} x_0 + \sqrt{1 - \bar{\alpha}_t} \, \epsilon_t, t \right) \right\|^2 \right]
\label{eq:simple_loss}
\end{equation}

Here, \(\epsilon_t \sim \mathcal{N}(0, I)\) denotes the standard Gaussian noise sampled at time step \(t\), and \(\epsilon_{\theta}\) is the noise prediction by the neural network. 

\clearpage
\renewcommand{\thefigure}{\thesection.\arabic{figure}}
\setcounter{figure}{0} 

\renewcommand{\thetable}{\thesection.\arabic{table}}
\setcounter{table}{0} 

\renewcommand{\theequation}{\thesection.\arabic{equation}}
\setcounter{equation}{0}
\section{Other Results}
\label{sec.quadrantB}

\subsection*{Dataset}

	\begin {figure}[h!]
	\centering{\includegraphics[scale=0.55]{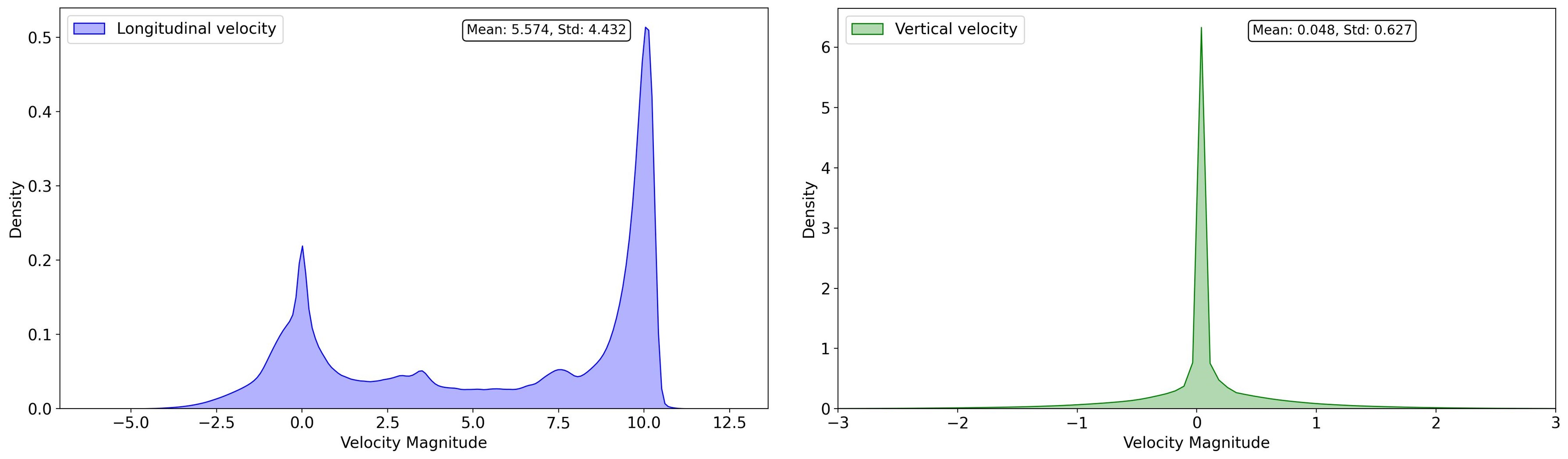}}
	\caption{Velocity distribution of longitudinal and vertical velocity component of CityFFD data. }
	\label{Figuredistribution}
	\end{figure} 

\subsection*{Reconstrucion performance of CAE}

The CAE model's reconstruction capability is a fundamental step toward ensuring its efficiency in capturing the turbulent flow field's essential dynamics and confirming its ability to predict the flow field's temporal evolution.
Consequently, this section examines the performance of the CAE model on the test dataset, using a latent vector dimension of 512.
To quantify the performance of CAE, the average of the reconstruction error, MSE, and the turbulent kinetic energy percentage, $E_k$, are reported in Table \ref{table.1}.
In previous studies, the percentage of turbulent kinetic energy has been commonly used as a reference for evaluating velocity fluctuations and their reconstruction \cite{MASOUMIVERKI2022104252, EIVAZI2022117038}, and is defined as follows:

\begin{equation}
E_k = \left( 1 - \left\langle \frac{\sum_{i=1}^{n}(w - \tilde{w})^2}{\sum_{i=1}^{n} w^2} \right\rangle \right) \times 100
\end{equation}

Here, $\langle \cdot \rangle$ represents time-based ensemble averaging, while $w$ and $\tilde{w}$ correspond to the reference velocity component and its reconstruction by the model, respectively, with $n$ denoting the number of grid points.

\begin{table}[h!]
\centering
\caption{The average reconstruction error, $MSE$, and the average of turbulent kinetic energy percentage, $E_k$, of the streamwise velocity components of the CAE models.}
\begin{tabular*}{\textwidth}{@{\extracolsep{\fill}} lcc}
\hline
 Model & $MSE$ & $E_{k}$  \\
\hline
CAE & 0.199  & 99.60\% \\

\hline
\end{tabular*}
\label{table.1}
\end{table}

The CAE shows strong performance in reconstructing the flow field. Due to the high magnitudes and variance of the streamwise velocity component, the CAE performs well in capturing the turbulent kinetic energy percentage. This suggests that the CAE effectively captures velocity fluctuations in the longitudinal direction.
\subsection*{Prediction performance of the DL models}

In the next step, the one-step prediction performance of CAE-LSTM (CAE with the parallel LSTM) and U-Net models is assessed to verify their effectiveness in the sequential time step predictions. 

Further, to quantify the one-prediction accuracy of different models, the average value of $mse$ of different models is listed in Table \ref{table.pred-error}.

\begin{table}[h!]
\centering
\caption{The average value of $mse$ of the reconstructed longitudinal component of one-step prediction of different models.}
\begin{tabular*}{\textwidth}{@{\extracolsep{\fill}} lccc}
\hline
&\multicolumn{3}{c}{Model}\\
\cline{2-4}
$mse$ & CAE  & U-Net & U-Net20 \\
\hline
$U$ component & 0.227  & 0.004 & 0.007 \\

\hline
\end{tabular*}
\label{table.pred-error}
\end{table}

It can be illustrated that the the velocity field reconstructed by different models is in good agreement with the CityFFD data. The skip connections in the U-Net architecture enhance its ability to capture local spatial information in turbulent flow fields, enabling the model to reconstruct complex flow dynamics with minimal absolute error.

\begin{figure}[h!]
    \centering
    
    \begin{minipage}[b]{0.3\textwidth}
        \centering
        \textbf{Time step = 6600}
    \end{minipage}
    \hfill
    \begin{minipage}[b]{0.3\textwidth}
        \centering
        \textbf{Time step = 7100}
    \end{minipage}
    \hfill
    \begin{minipage}[b]{0.3\textwidth}
        \centering
        \textbf{Time step = 7600}
    \end{minipage}

    \raisebox{-.5\height}{\rotatebox{90}{\textbf{CityFFD}}}
    \begin{minipage}[c]{0.3\textwidth}
        \centering
        \includegraphics[width=\linewidth]{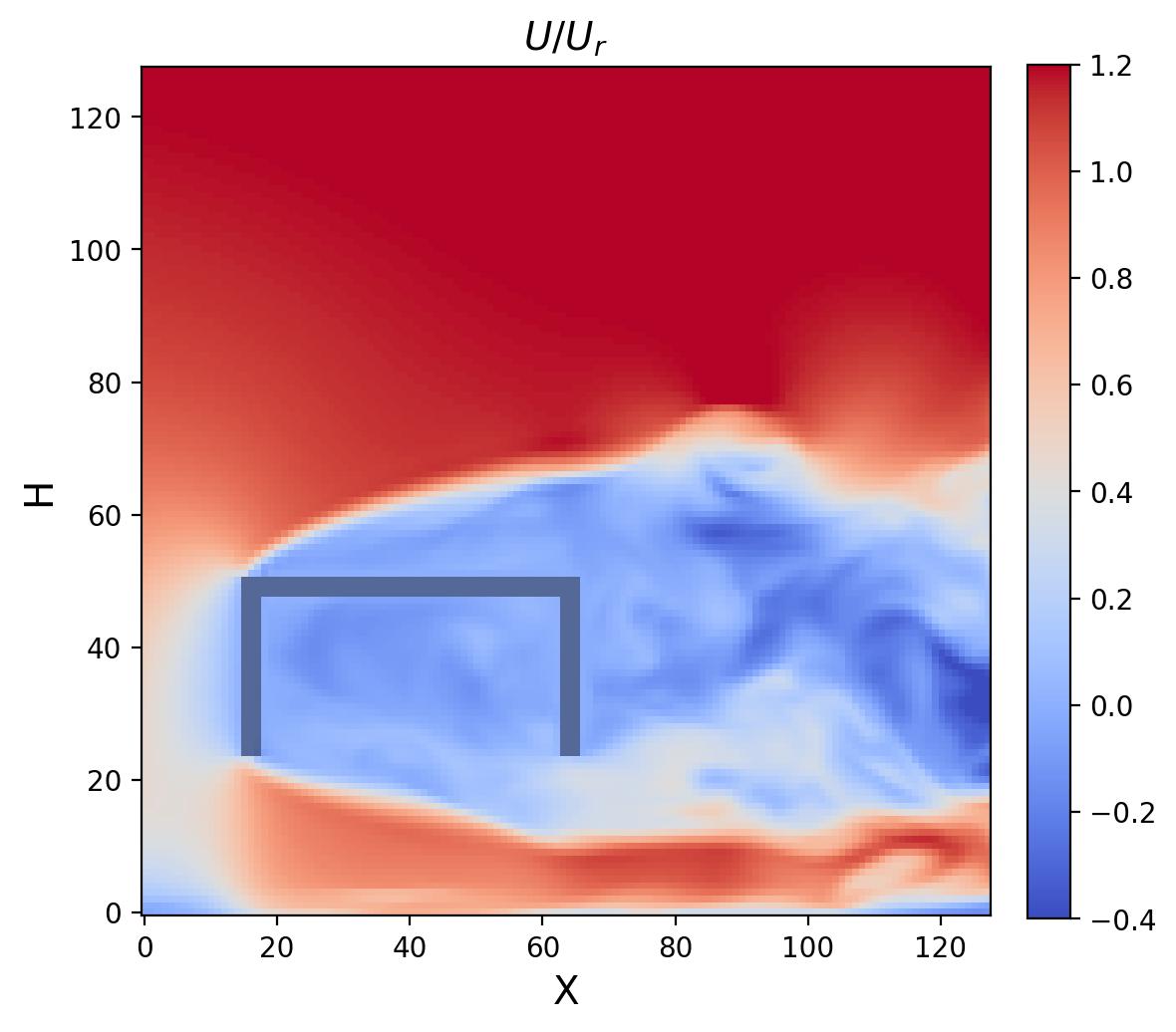}
    \end{minipage}
    \hfill
    \begin{minipage}[c]{0.3\textwidth}
        \centering
        \includegraphics[width=\linewidth]{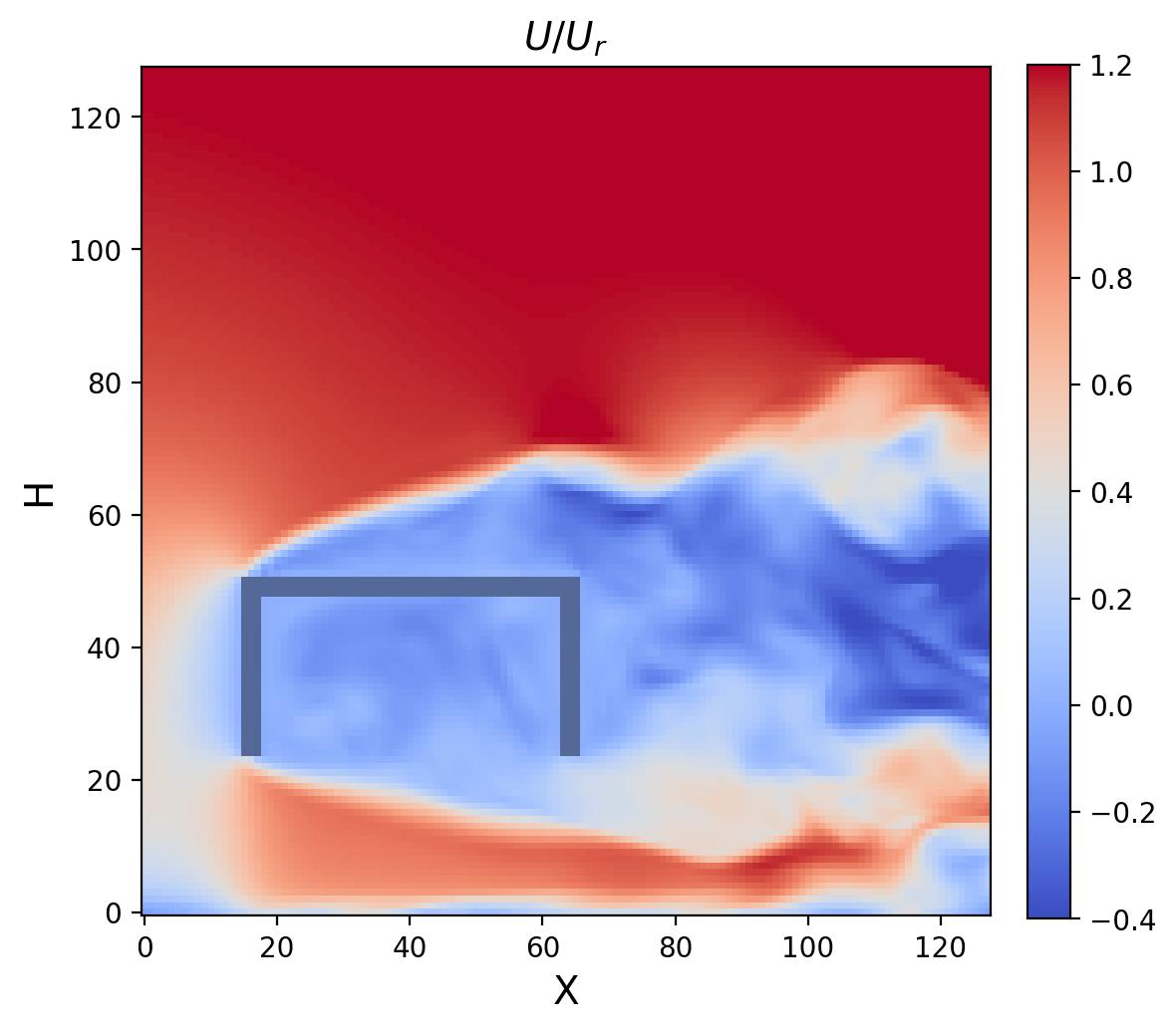}
    \end{minipage}
    \hfill
    \begin{minipage}[c]{0.3\textwidth}
        \centering
        \includegraphics[width=\linewidth]{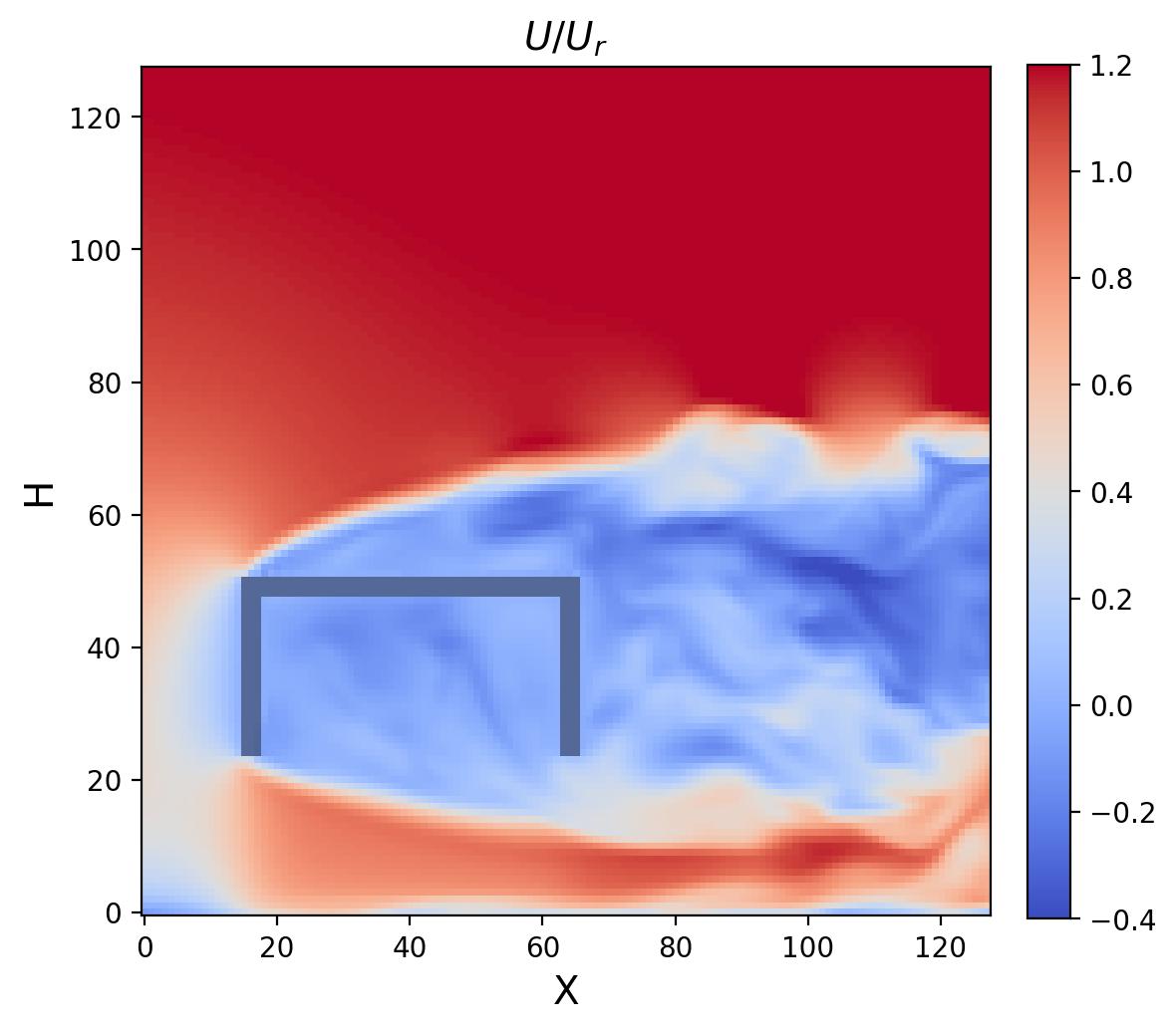}
    \end{minipage}

    \raisebox{-.5\height}{\rotatebox{90}{\textbf{CAE-LSTM}}} 
    \begin{minipage}[c]{0.3\textwidth}
        \centering
        \includegraphics[width=\linewidth]{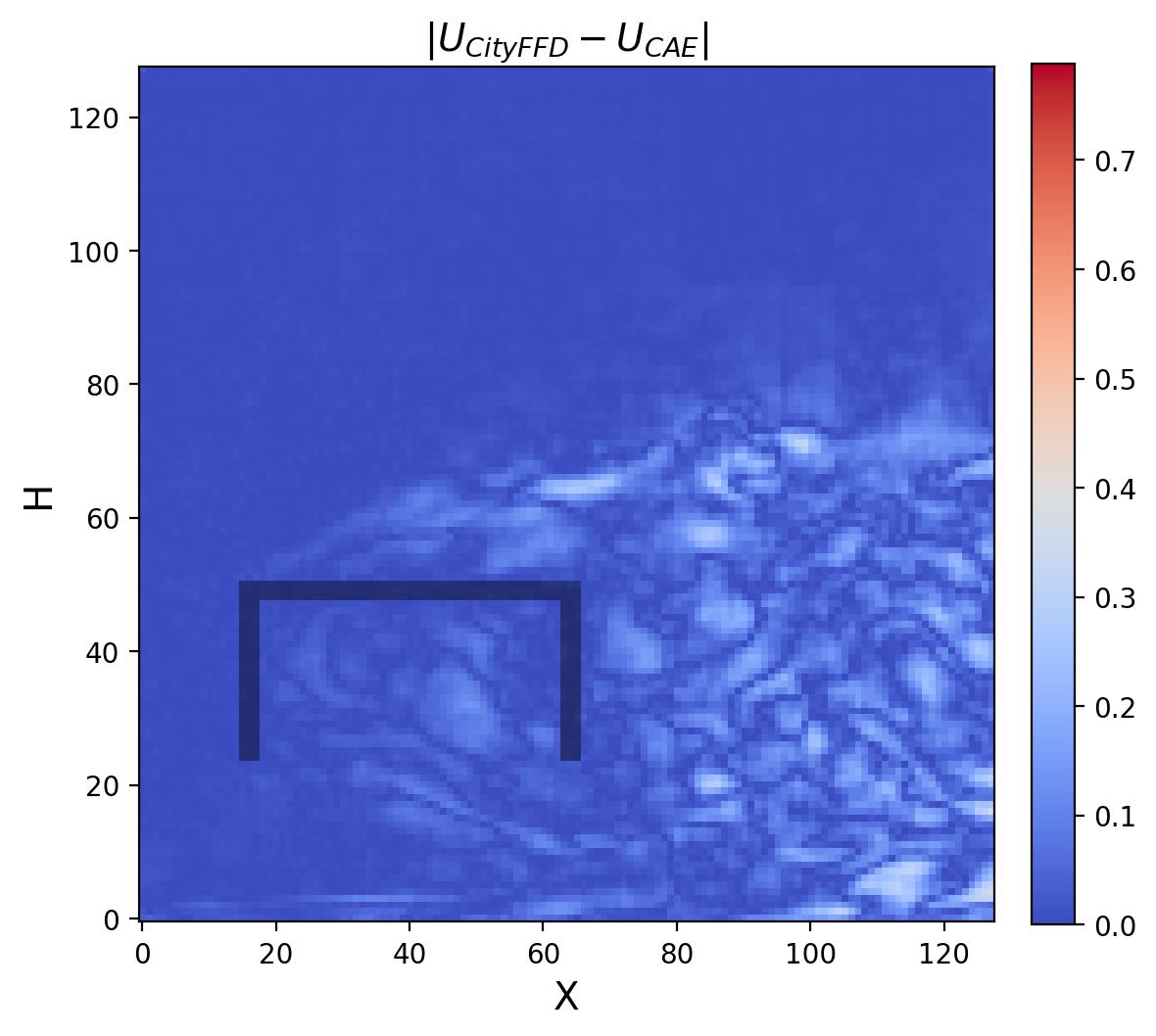}
    \end{minipage}
    \hfill
    \begin{minipage}[c]{0.3\textwidth}
        \centering
        \includegraphics[width=\linewidth]{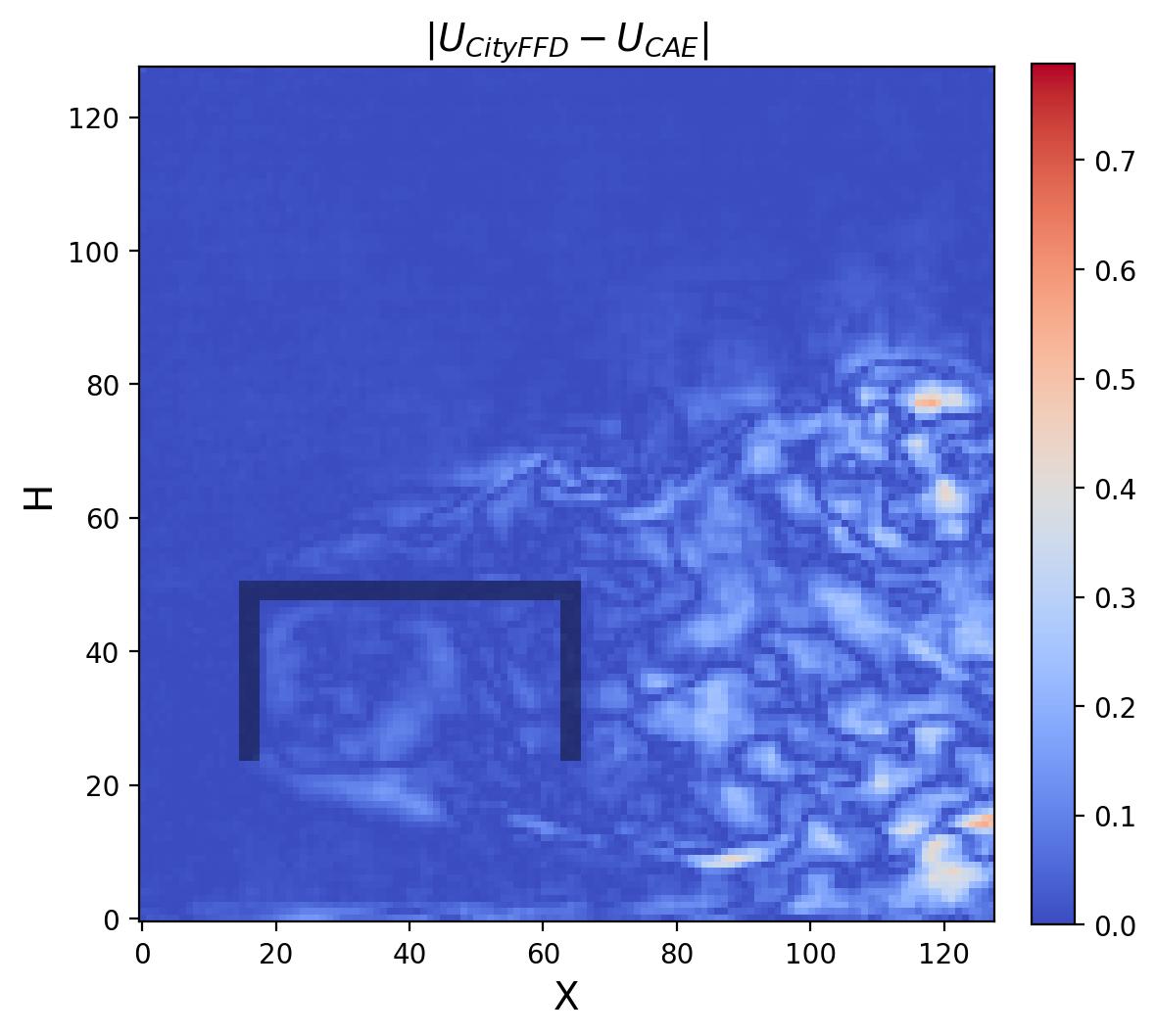}
    \end{minipage}
    \hfill
    \begin{minipage}[c]{0.3\textwidth}
        \centering
        \includegraphics[width=\linewidth]{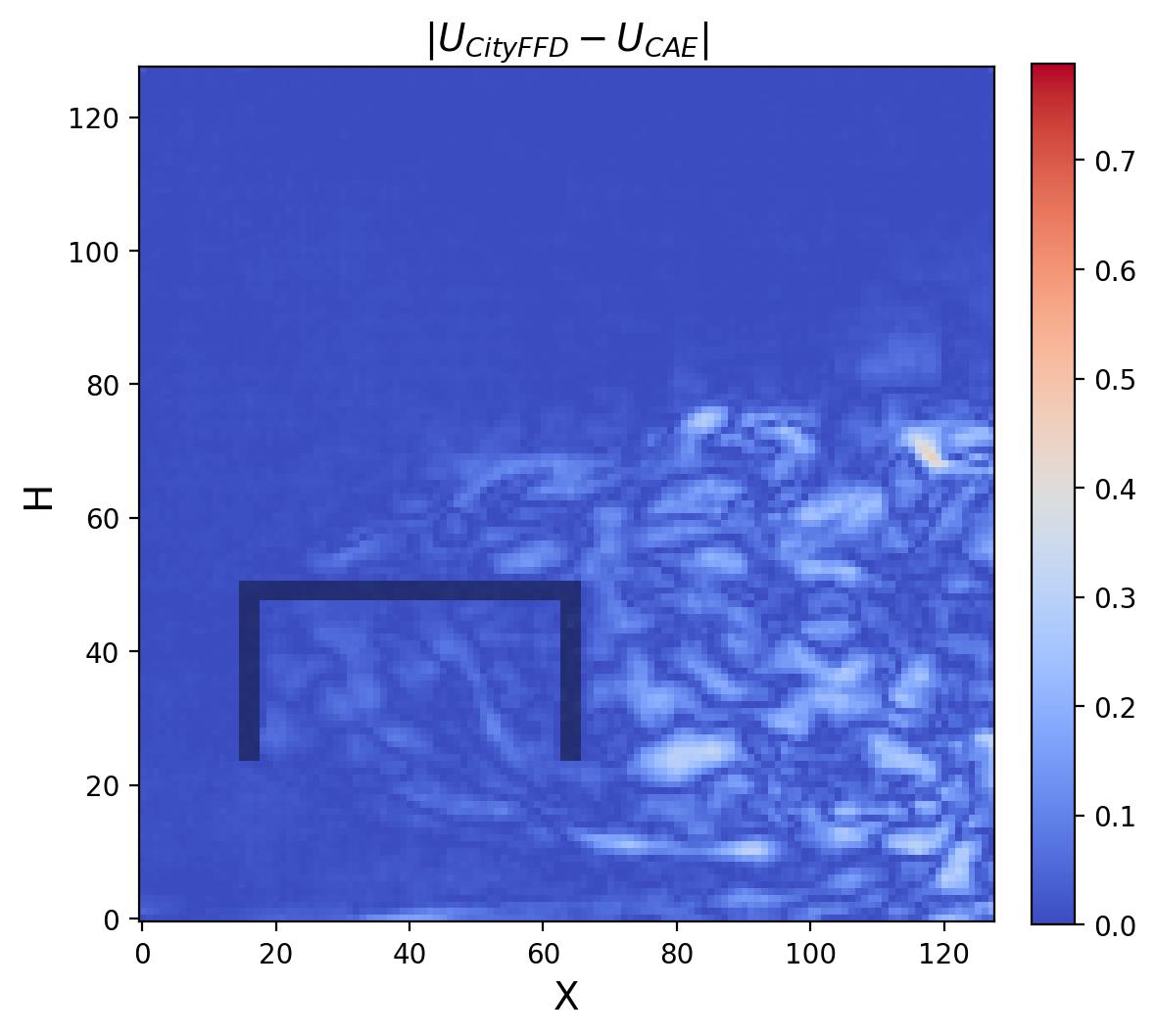}
    \end{minipage}

    \raisebox{-.5\height}{\rotatebox{90}{\textbf{U-Net}}} 
    \begin{minipage}[c]{0.3\textwidth}
        \centering
        \includegraphics[width=\linewidth]{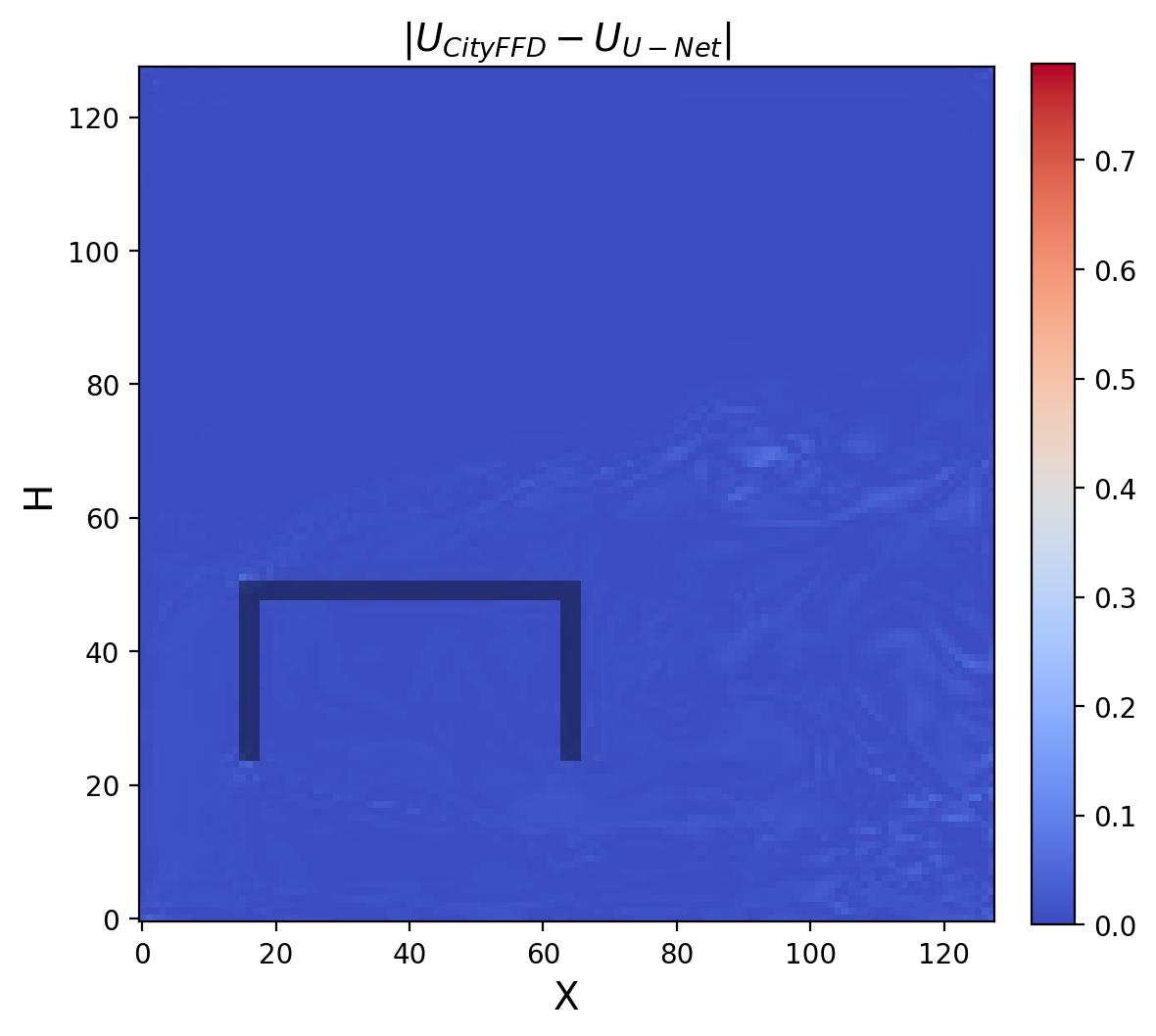}
    \end{minipage}
    \hfill
    \begin{minipage}[c]{0.3\textwidth}
        \centering
        \includegraphics[width=\linewidth]{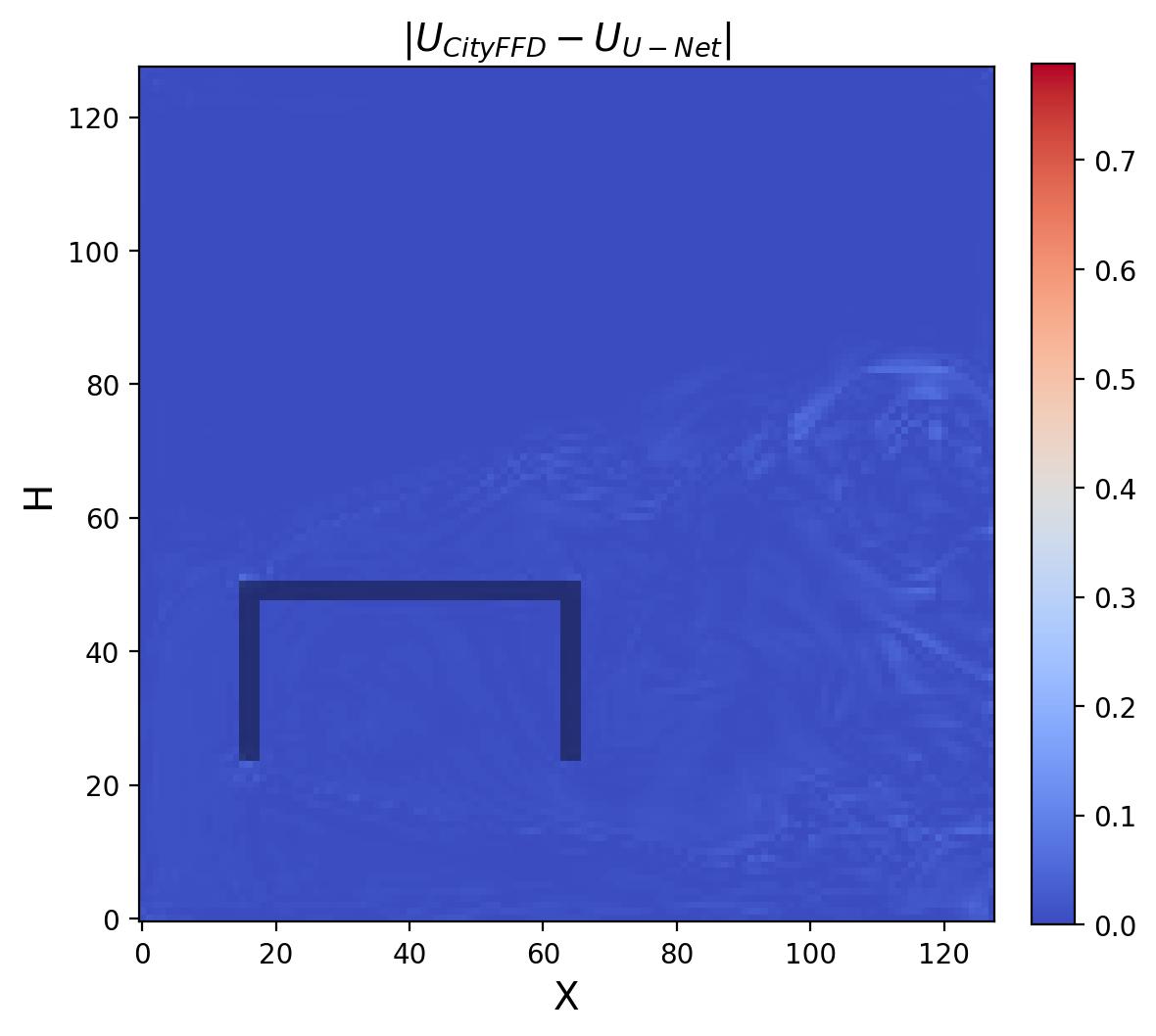}
    \end{minipage}
    \hfill
    \begin{minipage}[c]{0.3\textwidth}
        \centering
        \includegraphics[width=\linewidth]{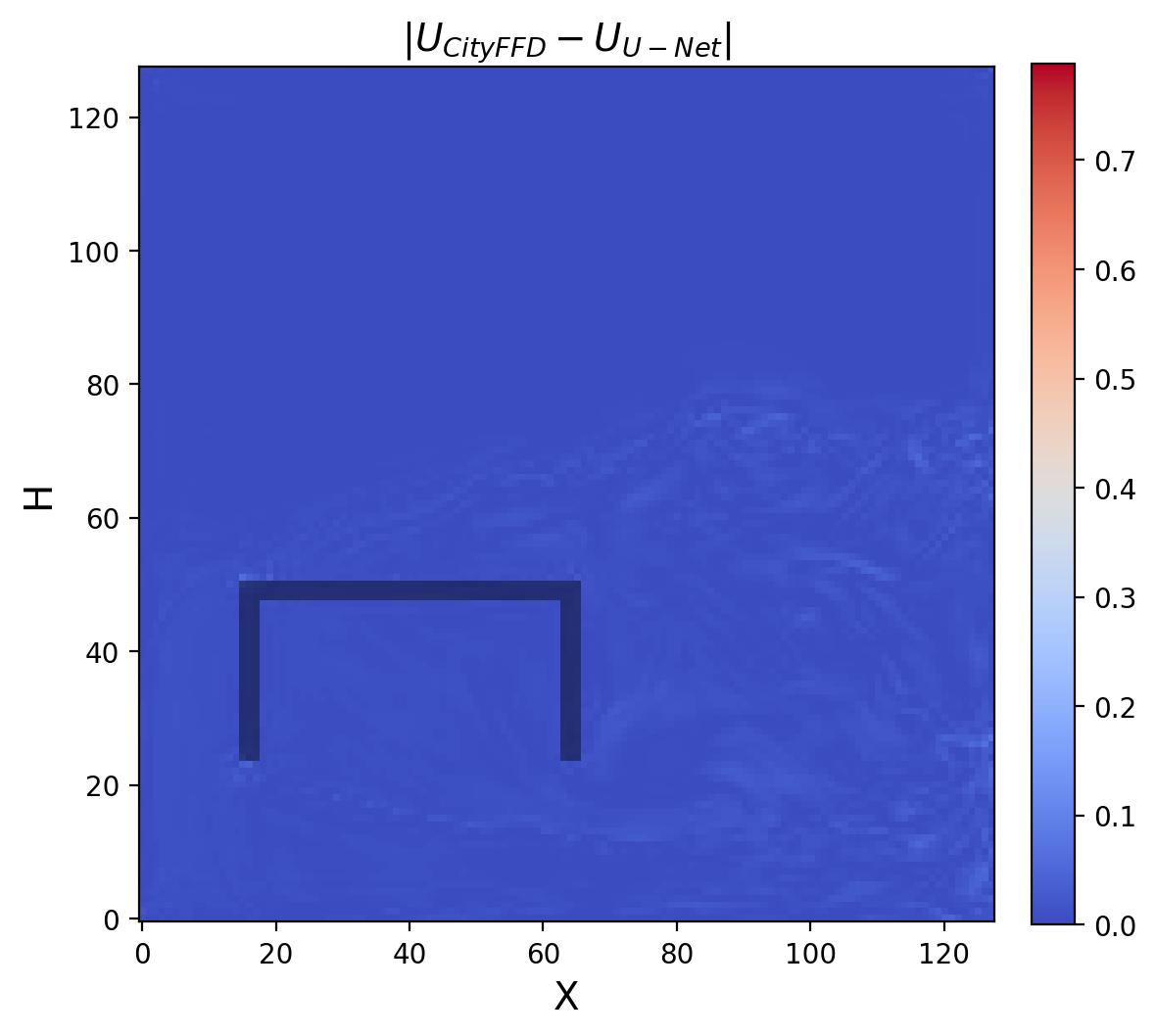}
    \end{minipage}

    \raisebox{-.5\height}{\rotatebox{90}{\textbf{U-Net20}}} 
    \begin{minipage}[c]{0.3\textwidth}
        \centering
        \includegraphics[width=\linewidth]{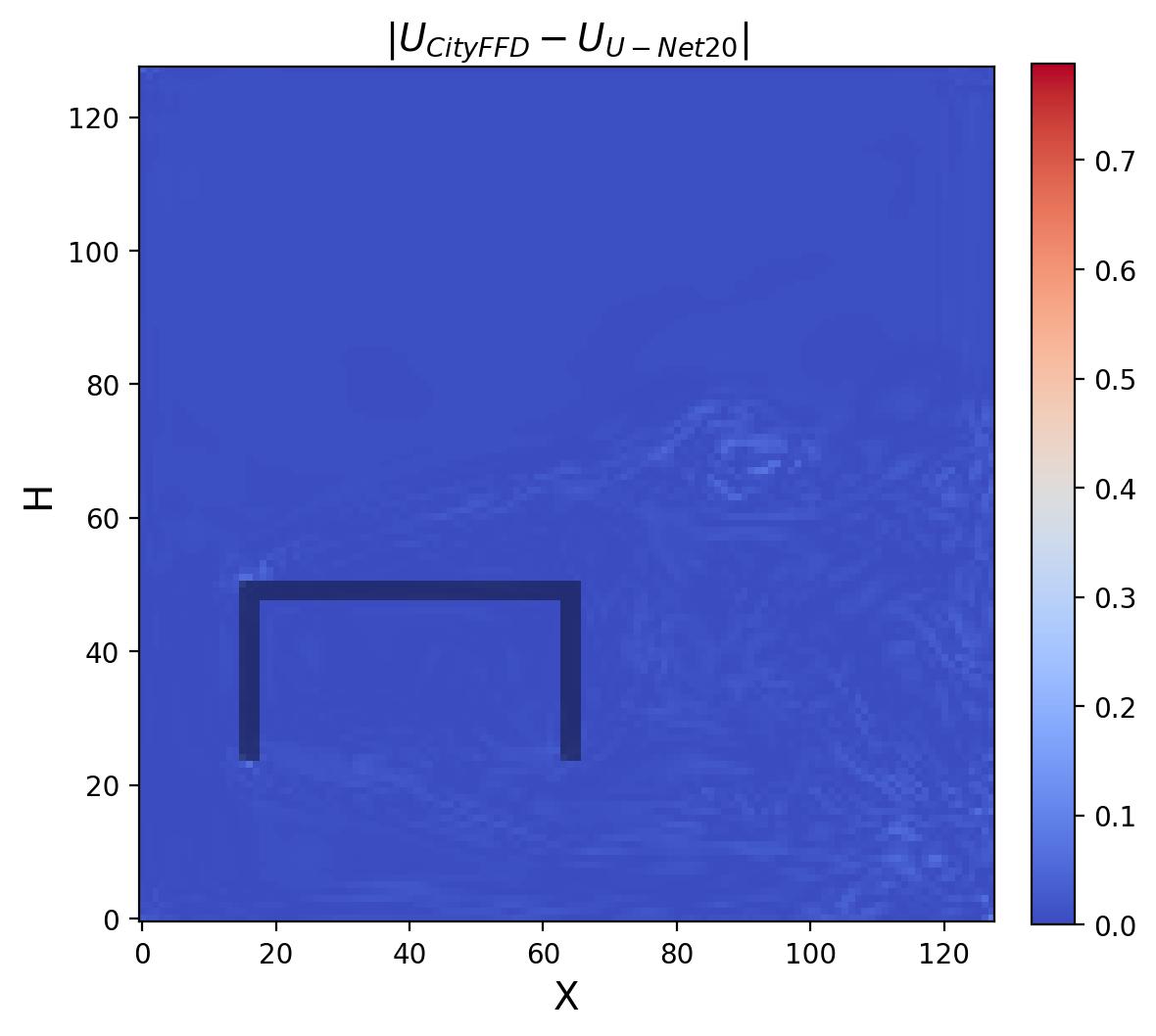}
    \end{minipage}
    \hfill
    \begin{minipage}[c]{0.3\textwidth}
        \centering
        \includegraphics[width=\linewidth]{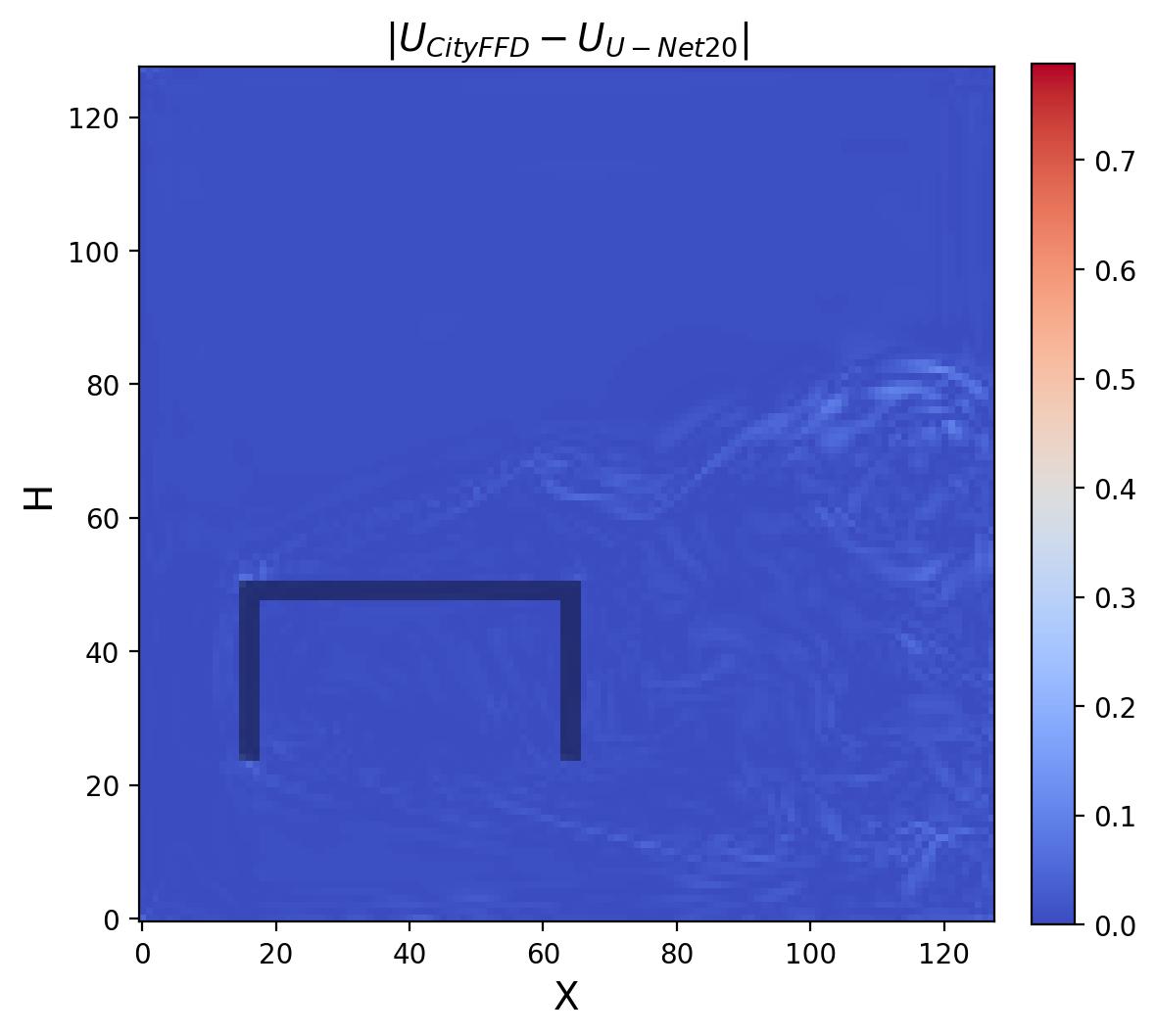}
    \end{minipage}
    \hfill
    \begin{minipage}[c]{0.3\textwidth}
        \centering
        \includegraphics[width=\linewidth]{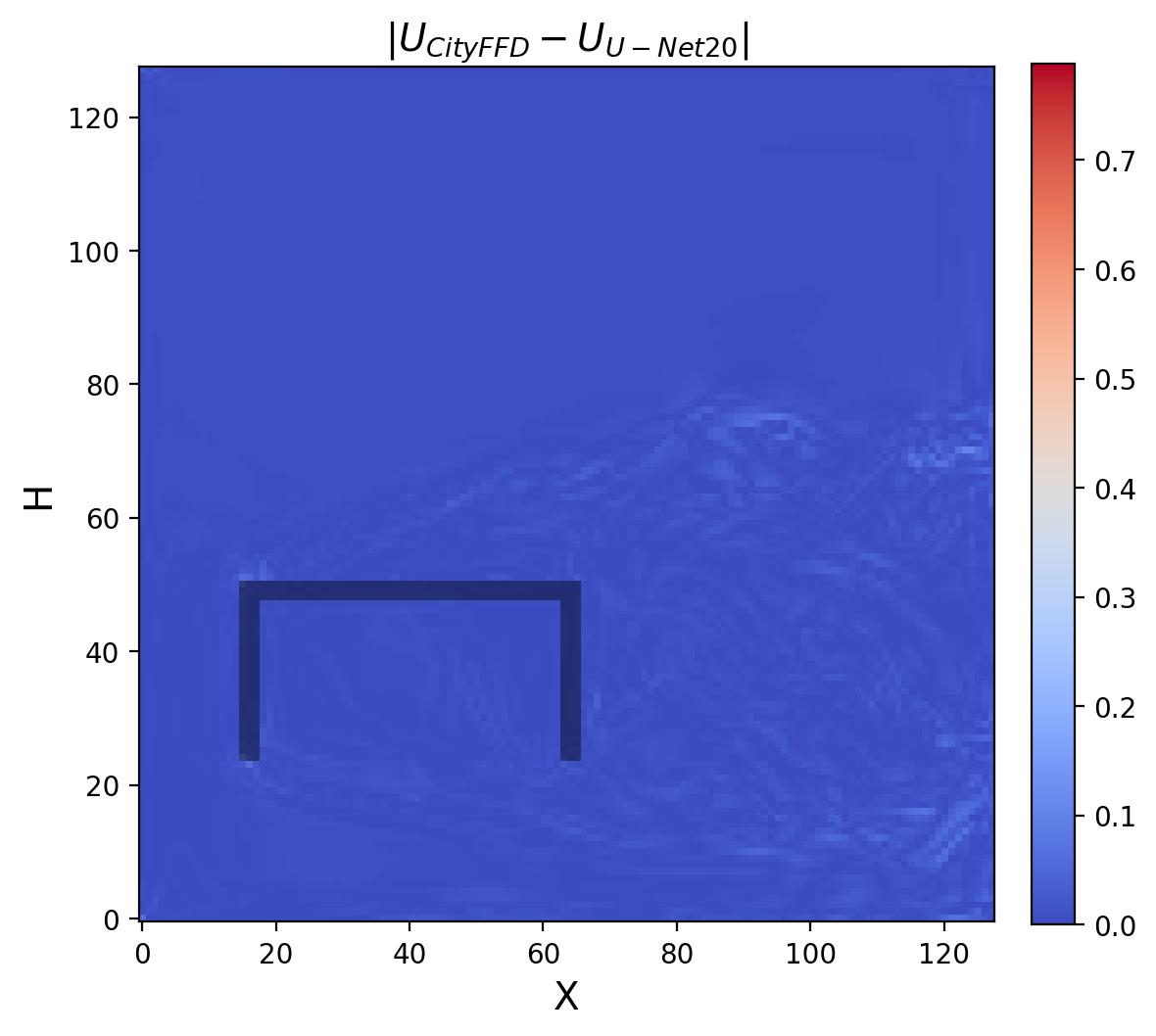}
    \end{minipage}

    \caption{The contour of absolute deviation of the dimensionless longitudinal velocity component reconstructed by CAE-LSTM, U-Neet, and U-Net20 against the CityFFD data at different time steps.}
    \label{fig.predict_u}
\end{figure}

To further visualize the prediction accuracy of different models, Figure \ref{fig.predict_u} depicts the contour of absolute deviation of the dimensionless longitudinal velocity reconstructed by different models against the ground truth at different time steps. 
It can be illustrated that the longitudinal velocity reconstructed by different models is in good agreement with the CityFFD data. The skip connections in the U-Net architecture enhance its ability to capture local spatial information in turbulent flow fields, enabling the model to reconstruct complex flow dynamics with minimal absolute error. In addition, due to the information loss in the bottleneck of CAE-LSTM, small deviations from ground truth are observed. However, all models demonstrated an acceptable accuracy and can be utilized for sequential time step predictions. 

\clearpage

\bibliography{main}
\bibliographystyle{unsrt}

\end{document}